\documentclass[a4paper,11pt]{article}

\usepackage{amsmath}
\usepackage{graphicx}
\usepackage{bm}

\begin{document}

\title{Controlling instabilities along a 3DVar analysis cycle by assimilating in
the unstable subspace: a comparison with the EnKF}

\author{A. Carrassi$^1$, A. Trevisan$^2$, L. Descamps$^3$, O. Talagrand$^4$ and F. Uboldi$^5$}
\maketitle

[1]: Institut Royal M\'et\'eorologique de Belgique, 1180 Bruxelles, Belgique$^*$

[2]: ISAC-CNR, 40129 Bologna, Italy

[3]: Laboratoire de M\'et\'eorologie Dynamique, \'Ecole Normale Sup\'erieure, Paris, France 

[4]: ISAC-CNR, 40129 Bologna, Italy

[5]: Novate Milanese, Italy\\

{\small $^*$Corresponding author e-mail: a.carrassi@oma.be}

%% The [] brackets identify the author to the corresponding affiliation, 1, 2, 3, etc. should be inserted.

%\runningtitle{Assimilating in the unstable subspace}

%\runningauthor{A. Carrassi et al.}

%\correspondence{A. Carrassi\\ (a.carrassi@oma.be)}

%\received{}
%\pubdiscuss{} %% only important for two-stage journals
%\revised{}
%\accepted{}
%\published{}

%% These dates will be inserted by the Publication Production Office during the typesetting process.

%\firstpage{1}

%\maketitle

\begin{abstract}

 A hybrid scheme obtained by combining 3DVar with the Assimilation in the Unstable Subspace (3DVar-AUS) is tested in a QG model, under perfect model conditions, with a fixed observational network, with and without observational noise. The AUS scheme, originally formulated to assimilate adaptive observations, is used here to assimilate the fixed observations that are found in the region of local maxima of BDAS vectors (Bred vectors subject to assimilation), while the remaining observations are assimilated by 3DVar. 
 The performance of the hybrid scheme is compared with that of 3DVar and of an EnKF. The improvement gained by 3DVar-AUS and the EnKF with respect to 3DVar alone is similar in the present model and observational configuration, while 3DVar-AUS outperforms the EnKF during the forecast stage. 
The 3DVar-AUS algorithm is easy to implement and the results obtained in the idealized conditions of this study encourage further investigation toward an implementation in more realistic contexts. 
\end{abstract}

\section{Introduction}

Data assimilation in meteorology and oceanography is experiencing a rapid phase of development, with flourishing of theoretical studies and applications. Traditionally the regular, globally available measurements network has been employed in the process of combining data with model forecasts to obtain the initial condition for model integrations. Recently there has been also a growing interest in {\it adaptive} observational systems (\cite{Langland2005}) and satellite data thinning (\cite{Ochotta_etal2005}).

While the merits of the relatively new ensemble methods are being compared with those of traditional variational methods, the potential benefit of the use of adaptive observations, added to the standard network, is being tested (\cite{Szunyogh02}).

The present is a follow-up study on the testing of an assimilation scheme developed by the authors (\cite{TU,UTC,UT,CTU}) and referred to as Assimilation in the Unstable Subspace (AUS). 
The basic paradigm of AUS is to track and control the analysis-forecast cycle instabilities. As a consequence, AUS has found its most natural applications in an adaptive observation context. Making use of just a few additional observations, properly located at each observing time in order to detect the analysis-forecast cycle instabilities estimated by Breeding on the Data Assimilation System (BDAS) it was possible to drastically reduce the forecast error.

Previous studies have in fact proven the AUS-BDAS system effectiveness in a hierarchy of dynamical systems characterized by different features and complexity, such as the Lorenz 40-variable model (\cite{Lorenz96}), the atmospheric quasigeostrophic model used here (\cite{RotBao96}) and a primitive equation ocean model (\cite{Bleck78}).
These results support, on the one hand, that BDAS, in view of its ability of tracking the instabilities of the assimilation system, provides a feasible targeting strategy; on the other hand that AUS is an efficient, computationally affordable, relatively easy to implement method to assimilate the adaptively located observations.
Moreover, they show that, even dealing with high-dimensional system, an efficient error control and an accurate state estimate can be obtained by monitoring only a reduced number of unstable directions and by properly using a limited number of observations.

By allowing ad hoc observations of the estimated instabilities, the deployment of an adaptive network naturally enhances the efficiency of AUS.
 In fact, targeting the spatial unstable structures improves the estimate of the amplitude of the unstable components of the error, thus enhancing the stabilizing effect. One natural question is therefore how AUS-BDAS performs with a fixed, standard, observational network.
As long as the spatial and temporal observational coverage of the standard network is dense enough, the instabilities can be detected as they move throughout the domain, although a reduced efficiency may be expected with respect to the adaptive observations case.

\cite{UT} investigated the use of fixed ``satellite'' surface elevation observations in a primitive equation oceanic model and pointed out the weakness of such an observational network. A stable error reduction could not be achieved by using only surface observations since dominating instabilities developed in deeper model layers before they could be detected at the surface. Instead, an efficient error reduction and a drastic improvement of the assimilation performance were obtained by means of vertical profiles adaptively located by BDAS and assimilated by AUS.

We investigate the best use of the AUS-BDAS method in an atmospheric context where only a network of fixed soundings is available. 

Errors, in an operational data assimilation cycle, are of various origin. Even if it is assumed that errors in the description of the system dynamics (the model error) are not present, noise is introduced at different stages of the observing and assimilation procedure.
It includes, among others, the measurement noise (coming also, for instance, by deficiencies in the retrieval procedures of indirect measurements), representativity errors and errors due to imperfections in the assimilation procedure itself.
Moreover, if the instabilities cannot be controlled by adequate (ad hoc) observations, nonlinearities become important. Thus, in the presence of large unstructured errors due to observation and system noise and to strong nonlinearities that can be expected to be important in a fixed observational setting, a failure of the unstable subspace paradigm of AUS is expected. These arguments suggest and motivate the combined use of AUS and a stationary assimilation method.

In the present study we propose an algorithm composed of a least-square based data assimilation scheme, the 3DVar, in combination with AUS, hereafter referred to as 3DVar-AUS. The added value of AUS over a 3DVar analysis cycle is investigated with a fixed observational network chosen to simulate a realistic observation coverage, and observation system simulation experiments are performed with both perfect and noisy observations.

Given the properties of the available observing network, a practical relevant question is how AUS compares with other assimilation methods.  For comparison, an ensemble Kalman filter (EnKF) (\cite{Even94,Even03}), has been implemented and optimized for the model and observational network used here. The EnKF is considered among the most promising ensemble-based sequential data assimilation schemes and therefore is an ideal candidate for the comparison with the proposed 3DVar-AUS scheme. 

This article is organized as follows. In Section 2 the details of the model and of the observational network are given. In Section 3 the three data assimilation schemes, objects of the comparison, are briefly described along with the specific implementation choices adopted here; particular emphasis is given to the proposed 3DVar-AUS. Results are described in Section 4, while the final summary and discussion can be found in Section 5.

\section{\label{Model} Model and observation network configuration   }

All the experiments described in this study are performed by making use of an atmospheric numerical model based on the quasigeostrophic equations in a periodic channel (\cite{RotBao96}). The model features mid-latitude large-scale flows and the channel is centered at 45$^o$N; its length is 16000 km while its width is 8000 km (approximately 180$^o$ of longitude by 70$^o$ of latitude). At the resolution of 250 Km used here, the model fields are discretized at 64 $\times$ 33 horizontal gridpoints on 5 vertical inner levels where potential vorticity, PV, is defined, plus 2 (top and bottom) levels where potential temperature, PT, is defined, so that the model state vector has dimension $I=14784$. The model solution is forced by relaxation to a zonal mean state with constant stratification and damped by a $\nabla^4$ horizontal diffusion and by an Ekman pumping at the surface. The time step for integration is approximately 30 minutes. 
At the resolution used, the model possesses 24 positive Lyapunov exponents, the leading one corresponds to a doubling time of 2.2 days. This model provides a qualitatively good representation of the true atmospheric mid-latitude dynamics while being simple enough to make long runs and statistical analysis feasible; it has been already used in some previous works on AUS (\cite{UTC,CTU}) as well as in a number of other data assimilation studies (\cite{Morss01,Corazza03,Corazza07}). An analysis of its dynamical properties can be found in \cite{snyderhamill03}.  

The forward model integration evolves from the analysis state at a given time $t_{k}$ to the forecast state at time $t_{k+1}=t_{k}+\tau$ where $k=0,1,2...$, and $\tau$ is the assimilation interval of a sequential assimilation scheme:
\begin{equation}
{\bf x}^{f}(t_{k+1})  = {\cal M} ({\bf x}^{a}(t_{k})),
\end{equation}
where ${\bf x}^{f}$ and ${\bf x}^{a}$ indicate the forecast and analysis states respectively.

The observing network consists of a fixed set of observations simulating synoptic profiles located at model gridpoints and measuring model variables, (potential vorticity, PV, at the five inner levels and potential temperature, PT, at top and bottom), at each level.

In a previous work with the same model (\cite{CTU}), AUS was used to assimilate a single adaptive observation in a large data-void area, and was combined with a 3DVar that assimilated all the fixed observations over a data-rich area.

\begin{figure*}[t]
\vspace*{2mm}
\begin{center}
\includegraphics[angle=270, width=8.3cm]{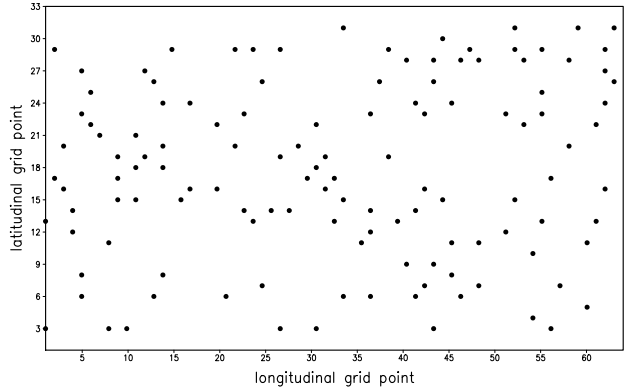}
\end{center}
\caption{\label{FIG1}The observational network used in the experiments. Each dot represents the horizontal location of a vertical sounding profile.}
\end{figure*}

In the present study, we intentionally work with a fixed network of observations to investigate the potential usefulness of AUS in the absence of adaptive observations that naturally enhance its efficiency. The observational network used here (Fig.~\ref{FIG1}) alternates data-rich/data-void areas as it is typical of a real observation network: the distribution of $M=125$ soundings is obtained through a random selection procedure.

Observing system simulation experiments are performed: a given trajectory, solution of the model equations, is taken to represent the true atmospheric evolution. At each analysis time, every $\tau=12$ hours, observations are taken by sampling the "true trajectory" at the observation locations. Following the widely used notation (\cite{ide97}), observation values are stored as components of the observation vector ${\bf y}^o$:
\begin{equation}
\label{obs-vect}
{\bf y}^o(t_k)  = {\cal H} ({\bf x}(t_k)) + \varepsilon(t_k)
\end{equation}
where $\varepsilon(t_k)$ represents the additive observation error, assumed to be white in time, Gaussian distributed with covariance ${\bf R}$ and ${\cal H}$ is the observation operator, estimating the observed variables from the model state.
Since the observations are located at model gridpoints and observe model variables, the observation operator is inherently linear and the notation ${\bf H}$ is used hereafter. 

Observation errors are assumed to be correlated in the vertical direction only and uncorrelated for different soundings.
The observational Root-Mean-Square (RMS) error at each level is set to $10\textrm{ }\%$ of the system's natural variability while, as in Morss (1999), the covariance between different levels is expressed in terms of the respective variances and of a vertical correlation function (Eq. 3.19 in \cite{bergman79}), depending on the vertical distance only.
 With these assumptions the observation error covariance matrix $\bf{R}$ takes the simple form of a diagonal block matrix, with a $7\times 7$ square matrix in each of the $M$ diagonal blocks. 

Following \cite{morss99} and  \cite{houtekamer93}, observation errors are randomly generated, and added to the true values, by sampling a Gaussian distribution $\mathcal{N}(0,{\bf R})$ with zero mean and consistent with the assumed observation error statistics.

\section{Data Assimilation Algorithms}

The analysis state at an arbitrary time $t_{k}$ is obtained as a linear combination of a forecast state, taken as a background field, with the observations:

\begin{equation}
\mathbf{x}^{a}=\left(\mathbf{I}-\mathbf{KH}\right)\mathbf{x}^{f}+\mathbf{Ky}^{o}=\mathbf{x}^{f}+\mathbf{Kd}\label{eq:ana}\end{equation}

where $\mathbf{d}=\mathbf{y}^{o}-\mathbf{Hx}^{f}$ is the innovation
vector, while the expression of the gain matrix $\mathbf{K}$
that minimizes the analysis error variance is the Kalman gain
 (\cite{Jaz70}):

\begin{equation}
\label{Kgain}
\mathbf{K}=\mathbf{P}^{f}\mathbf{H}^{\textrm{T}}\left[\mathbf{HP}^{f}\mathbf{H}^{\textrm{T}}+\mathbf{R}\right]^{-1}
\end{equation}

where $\mathbf{P}^{f}$ represents the forecast error covariance matrix.
In Eqs. (\ref{eq:ana}) and (\ref{Kgain}) all vectors and matrices refer to the same time $t_{k}$, and $T$ indicates matrix transposition.

\subsection{\label{3DVar} 3-Dimensional Variational Assimilation}

The three dimensional variational algorithm, 3DVar, used here has
been developed on the basis of that described in \cite{morss99}.
The forecast error covariance is estimated by the stationary matrix
$\mathbf{P}_{3DVar}^{f}=\mathbf{B}$ (background error covariance),
so that the analysis update is (see {\it e.g} \cite{talagrand97}):

\begin{equation}
\mathbf{x}^{a}=\mathbf{x}^{f}+\mathbf{BH}^{\textrm{T}}\left[\mathbf{HBH}^{\textrm{T}}+\mathbf{R}\right]^{-1}\mathbf{d}
\label{eq:3DVar}
\end{equation}

corresponding to the minimum of the 3D-Var objective function for
a linear observation operator $\mathbf{H}$. The matrix $\mathbf{B}$, that
is kept constant throughout the analysis cycle, was statistically
estimated by \cite{morss99} and has been multiplied here by
a scalar coefficient in order to optimize it 
for the present observational setting. 
The optimal value, chosen by minimizing the analysis error with the present network of 125 noisy profiles (Fig.1) is 0.97.

Equation (\ref{eq:3DVar})
is solved globally by using a conjugate residual solver algorithm (\cite{morss99}).
The initial conditions for all the experiments described in the text
is the final state of a 1-year-long 3DVar analysis cycle initialized
by randomly perturbing the truth state.

\subsection{\label{EnKF} Ensemble Kalman Filter}

The ensemble Kalman Filter, EnKF, used here is based on that described in Descamps and Talagrand (2007) and used for the study and intercomparison of ensemble initialization techniques. It basically follows the approach given by \cite{Even03, Even04}.

Let $\{{\bf x}^f_i(t_k)\}$ and $\{{\bf x}^a_i(t_k)\}$, with $i=1,2,..N$, be the ensemble of $N$ forecast and analysis states at times $t_k$, while $\bar{{\bf x}}^f(t_k)$ and $\bar{{\bf x}}^a(t_k)$ represent their respective means.
The forecast and analysis deviations from the mean, $\delta{\bf x}^{f,a}_i={\bf x}^{f,a}_i-\bar{{\bf x}}^{f,a}$ are stored as columns of the $I\times N$ matrices ${\bf X}^f$ and ${\bf X}^a$, respectively. The ensemble-based forecast and analysis error covariance matrices are then defined as:

\begin{equation}
\label{EnKF-Pf}
{\bf P}^f_{EnKF}  = \frac{1}{N-1}{\bf X}^f{\bf X}^{fT} \qquad {\bf P}^a_{EnKF}  = \frac{1}{N-1}{\bf X}^a{\bf X}^{aT}
\end{equation}

At each assimilation time $t_{k}$, a set of perturbed observations is generated, and then used for the ensemble analysis update, by randomly perturbing the actual observation vector:
\begin{equation}
\label{EnKF-obs}
{\bf y}^o_i  = {\bf y}^o+\varepsilon_i \qquad i=1,2,...,N,
\end{equation}
where $\varepsilon_i$ are independent realizations of the observational noise $\mathcal{N}(0,{\bf R})$.

At the analysis time the ensemble of forecast states is updated:

\begin{equation}
\label{EnKF-analysis}
{\bf x}^a_i = ({\bf I} - { \bf K}_{EnKF}{\bf H}){\bf x}^f_i + { \bf K}_{EnKF}{\bf y}^o_i \qquad i=1,2,...,N,
\end{equation}

where the gain matrix ${ \bf K}_{EnKF}$ is obtained by setting ${\bf P}^f={\bf P}^f_{EnKF}$ in Eq. (\ref{Kgain}). The ensemble of analysis states is then used to initialize $N$ full nonlinear model forecasts. 

In all the experiments described here, $N=30$ members are used and the very first analysis ensemble is generated by adding random noise to the reference initial condition. This noise is drawn from a Gaussian distribution with zero mean and variance equal to the average analysis error variance of the 3DVar experiment.

In order to avoid ensemble collapse, prevent filter divergence and optimize the EnKF performance, following \cite{DescampsTalagrand07}, forecast error covariance inflation (\cite{AndersonAnderson99}) and localization (\cite{houtekamer01}) have been adopted. The covariance inflation is obtained by multiplying the matrix ${\bf P}^f$ by a scalar coefficient before using it in the analysis:

\begin{equation}
\label{Pf-inflation}
{\bf P}^f_{EnKF} = (1+\alpha)\frac{1}{N-1}{\bf X}^f{\bf X}^{fT},
\end{equation}
where $\alpha$ is the (tunable) inflation factor.
As in Houtekamer and Mitchell (2001), the 5th order \cite{GaspariCohn99} function is used to localize the ${\bf P}^f_{EnKF}$,
the main parameter being the zero-correlation distance, $d_0$.

\begin{figure*}[t]
\vspace*{2mm}
\begin{center}
\includegraphics[width=8.3cm]{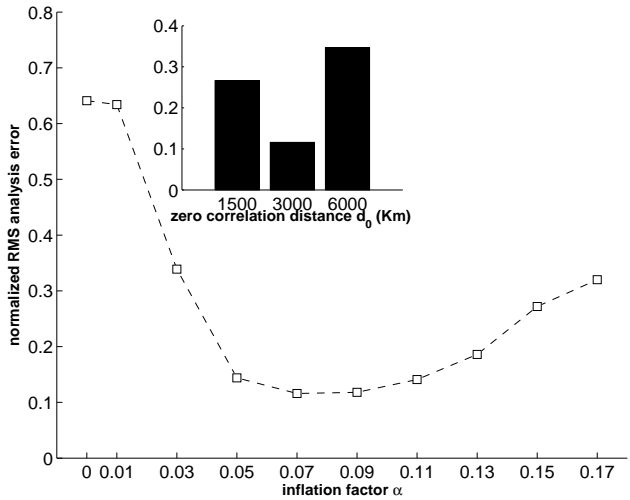}
\end{center}
\caption{\label{FIG2}Normalized time and domain RMS analysis error for the EnKF experiment as a function of the inflation factor ($\alpha$) by setting $d_0 = 3000$~Km, and, in the inset, as a function of the zero correlation distance $d_0$, by setting $\alpha = 0.07$. The experiments last 60 days and the average refers to the last 50 days. Errors are normalized with natural variability and expressed with potential enstrophy norm.}
\end{figure*}

Both $\alpha$ and $d_0$ have been optimized, over a 60 days analysis cycle, for the specific model and noisy observational network used here.
The best results are obtained with $d_0=3000$ Km and a $7\textrm{ }\%$ of inflation ($\alpha=0.07$). Figure \ref{FIG2} shows the normalized RMS analysis error as a function of the inflation factor (fixed zero correlation distance $d_0=3000$ Km), and as a function of the zero correlation distance (fixed inflation factor $\alpha=0.07$) in the inset.

\subsection{\label{AUS} Assimilation in the Unstable Subspace (AUS) combined with 3DVar}

\subsubsection{\label{AUS-back} Assimilation in the Unstable Subspace}

The mathematical formulation of AUS is briefly reported in the following, along with the description of the setup of its application in the present model and fixed observations configuration. A full description of its mathematical and theoretical foundations can be found in \cite{TU},\cite{UT} and \cite{CTU}.

The AUS method is basically aimed at tracking, and controlling, the instabilities of the (observationally forced) analysis-forecast cyclic system in view of their prominent role in the error evolution and their impact on the overall performance of the data assimilation scheme. To this end, when observations are available, an estimate of the analysis-forecast cycle unstable subspace is used to confine the analysis correction within this subspace, thus maximizing the effect of the assimilation where it is expected to be more necessary.

By combining Eqs. (1) and (\ref{eq:ana}), the evolution equation for the analysis state, the analysis-forecast cycle, is obtained:
\begin{equation}
\label{for-ana-cycle}
{\bf x}^a(t_{k+1})  = ({\bf I} - { \bf K}{\bf H}) {\cal M} ({\bf x}^a(t_{k})) + { \bf K}{\bf y}^o(t_{k+1}),
\end{equation}
Apart from the presence of a linearized observation operator ${\bf H}$, Eq. (\ref{for-ana-cycle}) is the governing equation of most sequential data assimilation schemes. 

The analysis-forecast cycle perturbation equations are given by: 
\begin{equation}
\label{BDAS-cycle}
\delta{\bf x}^a(t_{k+1})  = ({\bf I} - { \bf K}{\bf H}) {\bf M} \delta{\bf x}^a(t_{k})
\end{equation}
where ${\bf M}$, the Jacobian matrix of ${\cal M}$, represents the tangent linear model evolution. In principle, from Eq. (\ref{BDAS-cycle}), it is possible to estimate the unstable manifold of the forecast-analysis system. 

After storing the unstable vectors as the columns of a $I\times N$ matrix ${\bf E}$, and confining the analysis increment in the unstable subspace, the AUS analysis becomes:
\begin{equation}
\label{AUS-ana}
{\bf x}^a = {\bf x}^f + {\bf K}_{AUS}{\bf d}, 
\end{equation}
where:
\begin{equation}
\label{AUS-Ka}
{\bf K}_{AUS} = {\bf E\Gamma}({\bf HE})^T\left[{\bf R} + ({\bf HE}){\bf \Gamma}({\bf HE})^T\right]^{-1},
\end{equation}
or, equivalently:
\begin{equation}
\label{AUS-Kb}
{\bf K}_{AUS} = {\bf E}\left[({\bf HE})^T{\bf R}^{-1}({\bf HE})+{\bf \Gamma}^{-1}\right]^{-1}({\bf HE})^T{\bf R}^{-1},
\end{equation}
where, as before, all terms refer to the same time $t_k$ and where ${\bf \Gamma}$ is a $N\times N$ positive definite matrix, representing the forecast error covariance matrix in the unstable subspace:
\begin{equation}
\label{AUS-PF}
{\bf P}^f_{AUS} = {\bf E\Gamma E}^T.
\end{equation}

Suppose that a single  mode ${\bf e}$ is detected by a set of observations, then the AUS analysis reads: 
\begin{equation}
\label{AUS-ana-N1}
{\bf x}^a = {\bf x}^f + \frac{({\bf He})^T{\bf R}^{-1}{\bf d}}{({\bf He})^T{\bf R}^{-1}({\bf He})+\gamma^{-2}}{\bf e} 
\end{equation}

where $\gamma^{2}$, the forecast error variance in the ${\bf e}$ direction, can be estimated from innovations. 

From Eq.(\ref{AUS-ana-N1}) we see that the analysis increment vector has the direction of ${\bf e}$, and the amplitude that best-fits the observations. This means that, in physical space, the difference between the analysis and the forecast state has the three-dimensional structure of the unstable perturbation.

In practice, the unstable subspace of the data-assimilation system is estimated by means of an extension of the Breeding technique (\cite{TothKalnay93,TothKalnay97}), known as Breeding on the Data Assimilation System (BDAS) (\cite{TU}), which naturally incorporates the observational forcing in the perturbations dynamics.

Equation (\ref{BDAS-cycle}) is at the base of BDAS although in practice, as in the standard breeding, the full nonlinear model ${\cal M}$ is used instead of ${\bf M}$ to evolve initial random perturbations which are then kept small by repeatedly scaling them down to their initial amplitude.  
At each assimilation time, Eq. (\ref{AUS-ana}) is used to update both the control reference trajectory, solution of (1), and the set of the BDAS perturbed trajectories.  

This choice is motivated by the interest in testing and developing techniques feasible for realistic contexts where coding, implementing and keeping updated a tangent linear model may require some effort (for details on the implementation of BDAS see \cite{TU, UTC} and Sect. 3.3.2).

Furthermore, it is impractical to estimate the whole unstable subspace; this would require either a recursive orthonormalization to be applied to the set of BDAS vectors, according to the standard technique of \cite{Benetal80} or the use of computationally demanding techniques such as those described \cite{TrevPan98} and by \cite{WolfeSamel07}, which allow the estimation of the Lyapunov vectors.  
Instead, in this as in previous applications of AUS, a {\it refresh} procedure is introduced in the breeding cycle in order to systematically explore the unstable subspace of the system. In fact, as discussed in \cite{TU} and \cite{UT}, the more accurately the forecast error projection on a particular unstable direction is estimated, the more dominant will become the error projection on the complementary subspace.

Once a breeding cycle has been established, the BDAS modes and the forecast error will approximately have a structure given by a different linear combinations of independent unstable vectors of the underlying dynamics.

Local structures appearing in the global BDAS modes will be positively or negatively correlated with similar structures in the forecast error. It is therefore necessary to localize these structures before entering them in an assimilation expression of type (16). 
In addition, the localized structure needs to be observed, {\it i.e.} $\bf He$ should be non-zero and large enough to provide a reliable estimate of the analysis increment, whose sign is correctly determined by the innovation \footnote{By using $\mathbf{R}^{-1}$ as metrics, the norm of $\mathbf{He}$
is $\left(\mathbf{He}\right)^{\textrm{T}}\mathbf{R}^{-1}\left(\mathbf{He}\right)$: the components of $\mathbf{e}$
detected by different observations are weighted by the observational
accuracy. Noisy observations can then be effective if the $\mathbf{e}$
component they detect is large enough.}. 
It is important to notice that, by applying only a horizontal localizing function, the vertical structure of the perturbation is preserved in the definition of the analysis increment. 

While EnKF-like methods are based on an estimate of the forecast error covariance built from an ensemble of trajectories, the identification of unstable directions of the given control solution, which provide the three-dimensional structure of the analysis increment whose sign is determined by the innovation, is the distinctive feature of AUS.

EnKFs are based on the traditional Kalman filter equations and use a Monte Carlo approach to estimate the forecast error covariance matrix entering the analysis update, therefore they have to face the consequences of sampling errors with or without perturbed observations (\cite{WhitaHam02}) and filter divergence.
As a consequence, several ensemble filters formulations have been proposed to overcome these difficulties (for a review see \cite{Anderson03}, \cite{Even03}, \cite{Tippetetal2003}). Similarly, the AUS approach, that relies on the estimate of the forecast error projection on the unstable subspace, faces the practical difficulties of accurately estimating the unstable directions and the amplitude of the forecast error on each of them.
While the theoretical premises of the two approaches are different but clear, when it comes to the implementation all schemes are subject to several practical choices that make quantitative comparisons rather subjective.

An adaptive observations approach was followed by \cite{UT} and by \cite{CTU} in an oceanic and atmospheric model applications respectively.
In fact, the AUS approach of tracking and controlling the data assimilation system instabilities is most efficient when observations are indeed available at the moment and in the region where an instability develops. In other words, if the set of unstable vectors ${\bf E}$ can effectively be detected (that is to say if the components of ${\bf HE}$ are large enough), then ${\bf E}$ can be used in Eq. (\ref{AUS-ana}) to update the background state. When, as in the present application, only fixed synoptic observations are available, the efficiency of the AUS scheme is expected to be determined by the spacing and frequency of the observational network in relation to the typical spatial patterns and growing times of the system's instabilities.

\subsubsection{\label{3DVar-AUS} Implementation of 3DVar-AUS}

%Discutere il senso NATURALE dell'approccio ibrido per AUS con rete fissa
%PEZZETTI
Several approaches have been proposed to combine 3DVar with ensemble-based filters.
These hybrid Ensemble/3DVar analysis schemes introduce information on flow-dependent instabilities by properly weighting the static covariance with the ensemble estimated forecast error covariance (\cite{HamSny00,Wang_etal_07a}).
\cite{EtherBis04} investigated the robustness of hybrid schemes to model error: the model error modifies the optimal values of the weights to be given to the ensemble based and static covariance matrices, significantly reducing the weights corresponding to the former.

In the present study, we introduce a hybrid scheme that uses the flow-dependent covariance and the 3DVar covariance separately to assimilate different observations. The AUS assimilation is applied first to reduce the forecast error component in the model's unstable subspace estimated by the BDAS modes, using those observations that are able to detect them. The residual forecast error is assumed to be unstructured and a static 3DVar covariance is used to assimilate the remaining observations.
Possibly, error associated with unstable structures that were not present in the BDAS modes, or whose amplitude could not be reliably estimated with the available observations may be still present in the analysis and subsequent forecast: hopefully they will be accounted for at successive assimilation times.

The 3DVar-AUS forecast-assimilation and breeding cycle is implemented through the following recursive steps:

\begin{enumerate}
\item add $N$ ($=1$) random perturbations to the control analysis state (store the new set of perturbed states in place of those discarded; see step 9);
\item perform forecast integration of the control analysis and all perturbed states. The total number ($=12$) of perturbed trajectories evolved is equal to the {\it breeding time} (6 days) divided by the assimilation interval (12 hrs) times $N$ ($=1$);
\item determine the $N (=1)$ global BDAS modes to be used in the following step, as the difference between the perturbed trajectories that have completed a forecast-analysis breeding cycle (6 days) and the control forecast;
\item select from the $N (=1)$ global BDAS vectors the ($\hat{N}^o$, see below) local structures that are detected by observations;
\item estimate the forecast error variance, $\gamma^2$, from recent innovations;
\item apply AUS analysis, using the selected structures with relative observations, and making use of $\gamma^2$ (step 5), to the control and to the full set of ($=12$) perturbed trajectories;
\item store the innovations to be used in step 5;
\item apply 3DVar, using the remaining observations, to the control and to the full set of ($=12$) perturbed trajectories;
\item discard the trajectories relative to the $N (=1)$ BDAS modes used in steps 4-5, and rescale the remaining perturbations;
\item go back to step 1.
\end{enumerate}

The values given in parenthesis refer to the specific values adopted in the present application.

During the first 6 days of experiment, the control and all the perturbed trajectories are subject to only the 3DVar analysis update to assimilate all the observations; the AUS analysis cycle begins at day 7.

Thus, in all the experiments described below, a single $N=1$ BDAS mode is used at each assimilation time while the breeding time interval is set to 6 days. This latter value was chosen by tuning to minimize the average analysis error (see also \cite{CTU}). The breeding time is related to number and growth rate of the unstable directions of the reference trajectory, solution of the assimilation cycle. In practice the breeding time has to be long enough for the perturbations to acquire the structure of the local instabilities of the flow.

At the analysis times, instead of directly using the global BDAS modes ${\bf e}$ in Eq. (16), $\hat{N}$ local isolated structures are extracted from ${\bf e}$. Once the unstable structures are selected, we identify the observations, if any, able to detect the $\hat{N}$ local structures; this defines a set of $\hat{N}^{o}\le \hat{N}$ $"$observable$"$ structures which are finally used in the analysis update. Details on steps 4-6 are given in the following: 
\begin{itemize}
\item {\bf (a)}
Selection and localizations of structures. The localization is made by repeating the point-by-point multiplication of an horizontal Gaussian function, $f({\bf x})=e^{\vert\vert{\bf x}-{\bf x}_{max}\vert\vert^2/d^2}$ ($d$ being the characteristic distance), centered on the local maxima, ${\bf x}_{max}$, of each BDAS mode ${\bf e}$. The process starts from the absolute maxima; subsequent maxima are searched at a distance larger than twice $d$. The $\hat{N}$ extracted structures are therefore assumed to be uncorrelated. We point out that the localization is made in the horizontal direction only, so that the vertical structure of the local maxima is preserved. 
\item {\bf (b)}
Identify available observations and observable structures. The observations selection is made in two steps: 
all the structures with no observations within an horizontal $\left(l,l\right)$ gridpoints box, centered on their maximum, are discarded; 
the amplitude of the structure at the observation locations must be larger than a fixed ratio $\beta$ of its maximum amplitude. At the end of these steps, $\hat{N}^{o}\le \hat{N}$ $"$observable$"$ structures, $\hat{{\bf e}}_i^{o}$, are obtained, each one detected by $M_i$ observations at the same level, $i=1,...,\hat{N}^{o}$. For each of the $\hat{N}^{o}$ observable structures, define one innovation vector, ${\bf d}_i$, one observation error covariance, the $M_i \times M_i$ matrix ${\bf R}_i$, one observation operator, the $M_i \times I$ matrix ${\bf H}_i $ and one scalar $\gamma_{i}^{2}$. 
\item {\bf (c)} 
AUS analysis update. If $\hat{N}^{o}\ne0$, that is at least one observable structure $\hat{{\bf e}}_i^{o}$ is identified, the AUS analysis update is performed, otherwise all the observations are assimilated with 3DVar. The analysis is made according to Eq. (16), using the $\hat{N}^{o}$ structures $\hat{{\bf e}}_i^{o}$ sequentially: 

\begin{equation} 
\label{AUS-ana-N1-2} 
{\bf x}^a = {\bf x}^f + \sum_{i=1}^{\hat{N}^{o}}\frac{({\bf H}_i\hat{{\bf e}}_i^{o})^T({\bf R}_i)^{-1}{\bf d}_i}{({\bf H}_i\hat{{\bf e}}_i^{o})^T({\bf R}_i)^{-1}({\bf H}_i\hat{{\bf e}}_i^{o})+(\gamma_i)^{-2}}\hat{{\bf e}}_i^{o} 
\end{equation}

Expression (\ref{AUS-ana-N1-2}) is the AUS analysis equation in the case of $\hat{N}^{o}$ isolated structures each one observed by $M_i$ observations whose associated error covariance matrices are ${\bf R}_i$. 

Given the characteristics of the observational network used here and in view of the selection procedure (point (b))
, each set of $M_i$ observations associated to the observable structure $\hat{{\bf e}}_i^{o}$ consists of scalar measurements at the same level (where the structure maximum is located). As a consequence, the observations are uncorrelated and all have the same variance $\sigma^2$: ${\bf R}_i=\sigma^2{\bf I}_i$, where ${\bf I}_i$ is the identity matrix of order $M_i$. By making use of these hypotheses, after rearranging, Eq. (\ref{AUS-ana-N1-2}) becomes:

\begin{equation}
\label{AUS-ana-N1-3}
{\bf x}^a = {\bf x}^f + \sum_{i=1}^{\hat{N}^{o}}\frac{\gamma_i^2({\bf H}_i\hat{{\bf e}}_i^{o})^T({\bf H}_i\hat{{\bf e}}_i^{o})}{\sigma^2+\gamma_i^2({\bf H}_i\hat{{\bf e}}_i^{o})^T({\bf H}_i\hat{{\bf e}}_i^{o})}\frac{({\bf H}_i\hat{{\bf e}}_i^{o})^T{\bf d}_i}{({\bf H}_i\hat{{\bf e}}_i^{o})^T({\bf H}_i\hat{{\bf e}}_i^{o})}\hat{{\bf e}}_i^{o} 
\end{equation}
Equation (\ref{AUS-ana-N1-3}) is the AUS analysis update used in the experiments described hereafter. Each of the scalars $\gamma_i^2$, representing the variance of the forecast error along the structure $\hat{{\bf e}}_i^{o}$, needs to be estimated.

In practice the terms $\gamma_i^2({\bf H}_i\hat{{\bf e}}_i^{o})^T({\bf H}_i\hat{{\bf e}}_i^{o})$ are estimated statistically using the innovations, according to the approach introduced in \cite{CTU}: details can be found in the Appendix.
\end{itemize}
Once the AUS analysis is completed, the analysis field is used as the background in the 3DVar analysis to assimilate the remaining $M-\sum_{i=1}^{\hat{N}^{o}}M_i$ observations (step 8).\\

As mentioned above, a refresh procedure (steps 1 and 9) is used in the implementation of BDAS (\cite{TU}). 
At each assimilation time, after the analysis update, the set of BDAS modes is discarded after being used in the assimilation. A new one is introduced, which starts to undergo a new breeding procedure: after completion of its breeding period it will be used in an assimilation step. Although this procedure increases the computational cost of BDAS (with respect to evolving the same set of perturbed states), it is very beneficial to efficiently span the trajectory's unstable subspace and to provide a reliable estimate of the data assimilation system instabilities. The refresh was also used in \cite{CTU} in the context of the same QG model used here, while a trade-off procedure between keeping and discarding the $N$(=6 in that case) BDAS modes altogether was used in \cite{UT} in the context of a primitive equation ocean model. 

In the experiments described here, the characteristic distance is chosen, after optimization, to be $d=1500$ Km. Also, a limit to the total number of the extracted assimilating structures is set, $\hat{N}=20$. Furthermore, the size of the searching box $l$ and the coefficient $\beta$ have been optimized by tuning, and are equal to $l=7$ gridpoints (equivalent to an area of 1500 Km$^2$) and $\beta=0.6$

A practical aspect which has to be taken into account when comparing 3DVar-AUS and EnKF is the computational cost. Both schemes, the hybrid 3DVar-AUS and the EnKF, require high computational power determined mainly by the number of evolving trajectories (perturbations of control for the 3DVar-AUS; ensemble members for the EnKF). At the analysis step, the computational most demanding part in the EnKF are the large matrix multiplication and inversion needed to compute the gain matrix; in 3DVar-AUS, the computational cost is mainly due to the 3DVar analysis update which has to be repeated for all the simultaneously evolved perturbed trajectories. A detailed comparison of the computational cost of the two algorithms needs to be done case by case, since the implementation choices are expected to depend on the system under consideration and on the available observation network. The computational cost, in terms of computing time, in the present application are given at the end of the next section.

\section{\label{Results}Results}

\subsection{Experiments with perfect observations}

We first compare the performance of the algorithms in a perfect observation setting.

Clearly real observations are not perfect. Anyhow, the idealized perfect observations setting is related to the problem of observability of the assimilation system and provides insight on the ability of various methods to track the instabilities of the system. The use of perfect observations with AUS allows for an accurate evaluation of the analysis increment and potentially reduces to zero the analysis error projection along the unstable structures used in the analysis update (18); the relation between the observability condition and AUS was discussed by \cite{TU}. Furthermore, the perturbed observation case provides an upper limit of performance of the assimilation schemes under comparison.

Perfect observations are obtained by applying the observation operator to the true state; $\varepsilon=0$ in Eq. (\ref{obs-vect}) and $\varepsilon_i=0$ in Eq. (\ref{EnKF-obs}). The observation error covariance matrices are set to zero in Eqs. (\ref{eq:3DVar}) and (\ref{EnKF-analysis}) and $\sigma^2=0$ in (\ref{AUS-ana-N1-3}). Under these conditions all EnKF members assimilate the same (unperturbed) observations ${\bf y}^{o}$; the parameters $\alpha$ and $d_0$ of the EnKF experiments are set to the same values optimized in the noisy case, $\alpha=0.07$ and $d_0=3000$ Km.

\begin{figure*}[t]
\vspace*{2mm}
\begin{center}
\includegraphics[width=8.3cm]{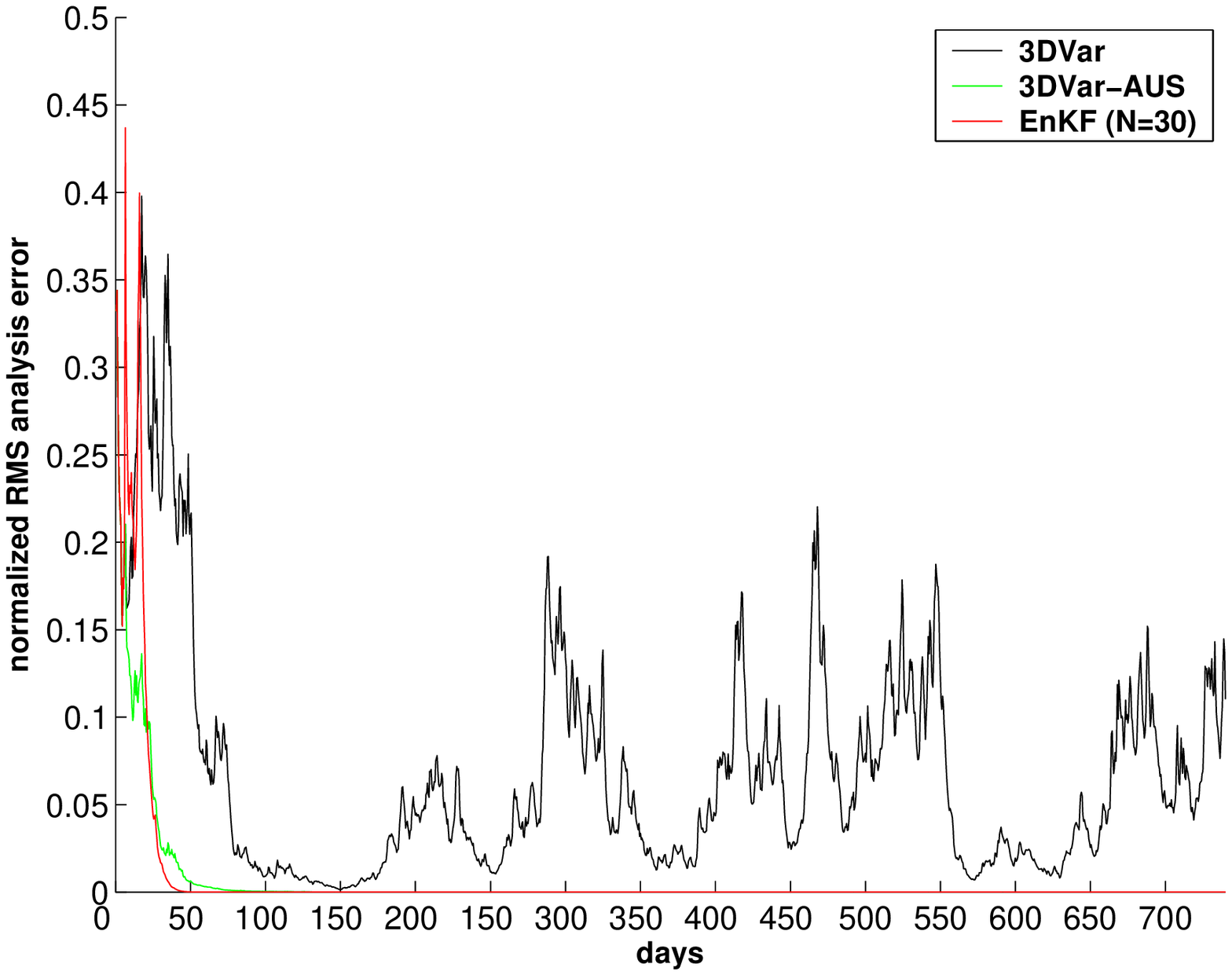}
\end{center}
\caption{\label{ERR-perf} Perfect observations: normalized RMS analysis error as a function of time for 3DVar (black), 3DVar-AUS (green) and the EnKF (red). Errors are normalized with natural variability and expressed with potential enstrophy norm.} 
\end{figure*}

Figure \ref{ERR-perf} shows the normalized RMS analysis error, as a function of time, for 3DVar, 3DVar-AUS and EnKF over 2 years. The 3DVar analysis error undergoes fluctuations with large spikes, presumably related to an unreliable representation of the actual forecast error in the stationary estimate ${\bf B}$ of the forecast error covariance matrix.

The average RMS analysis error, excluding the first 100 days, is about $6\textrm{ }\%$ of the system's natural variability. The behaviour of 3DVar-AUS and EnKF is impressive: the average RMS analysis error, excluding the first 100 days, is about
$1.6\times 10^{-6}$ ($0.00016\textrm{ }\%$) and $5.0\times 10^{-7}$ ($0.00005\textrm{ }\%$)
 of the system's natural variability, respectively. At the very beginning of the assimilation cycle, the EnKF solution drifts away from the truth, probably because the ${\bf P}^f_{EnKF}$ has not yet acquired the dynamic consistency, then, within about 50 days, the analysis error rapidly decreases and remains confined to very low values for all the subsequent period. 
Analogously the 3DVar-AUS experiment takes a transient time before error reduction; after about 80 days the 3DVar-AUS analysis error drops to very low values until the end of the experiment.   

\begin{figure*}[t]
\vspace*{2mm}
\begin{center}
\includegraphics[angle=270, width=18cm]{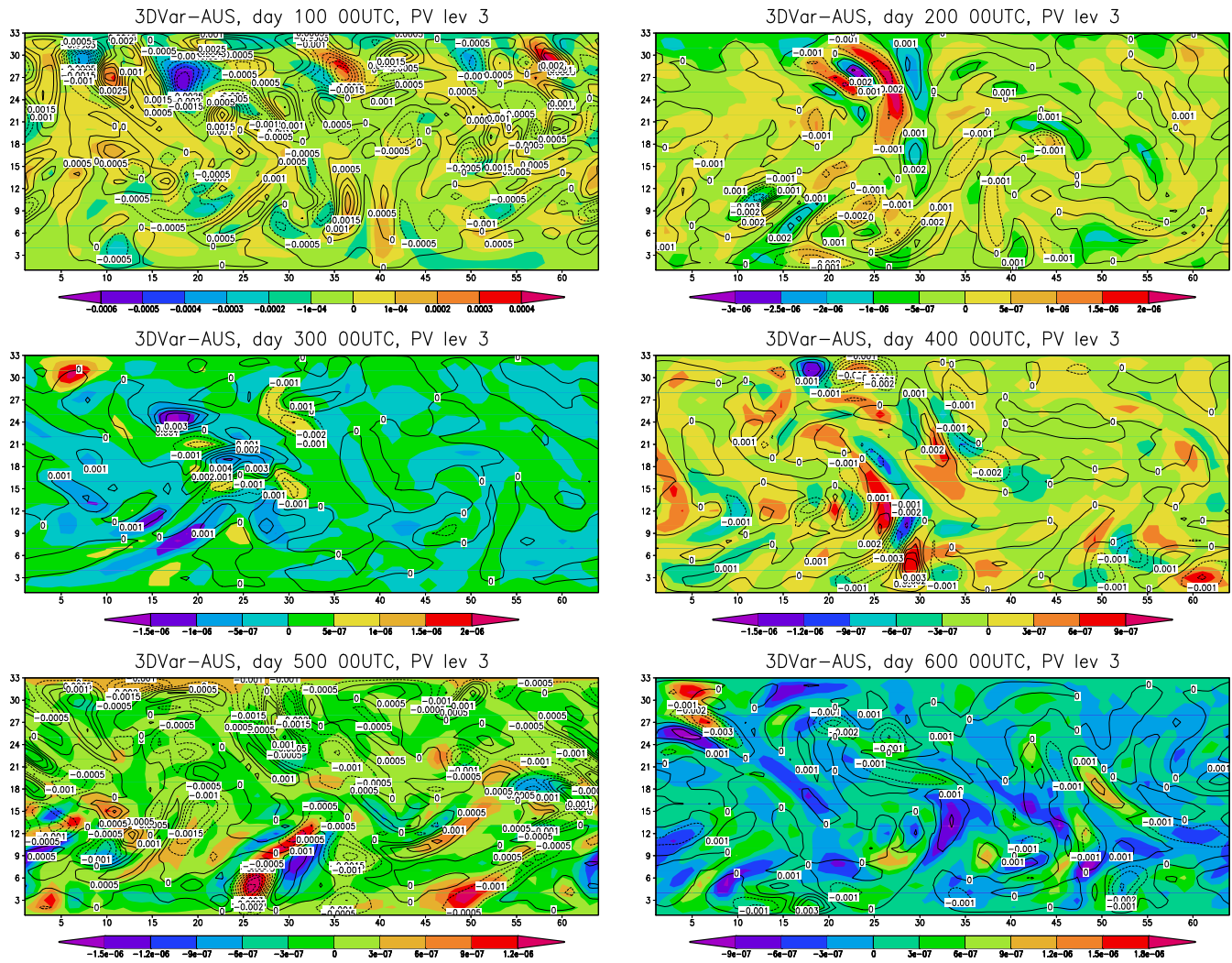}
\end{center}
\caption{\label{BDAS-perf}Perfect observations: mid-level PV background error (shaded) and PV bred mode (contour) at days: 100, 200, 300, 400, 500, 600 along the perfect observation 3DVar-AUS experiment.}
\end{figure*}

Figure \ref{BDAS-perf} shows the current BDAS mode superimposed to the actual forecast error, at 6 instants during the 3DVar-AUS perfect observations experiment.
Although the coincidence is not systematic, the BDAS mode ability to capture the relevant components of the forecast error appears evident.
The actual forecast error patterns appear composed of spatially distributed structures, often in a dipole or quadrupole configuration, separated by wide, relatively homogeneous, low error regions. 
Such pronounced and localized structures, presumably related to the instabilities of the underlying dynamics, are also frequently found in the corresponding BDAS mode. 
These are the instabilities we want to track and control. If they are not eliminated by the assimilation at one analysis time, they may still migrate to observed areas and, possibly, be detected at a later time. 
From Fig.~\ref{BDAS-perf} we also see that the spatial scale of these structures does not exceed 3000 Km (11 gridpoints): this supports the use of the Gaussian masking function with $d=1500$ Km.

The local character of the atmospheric instabilities are described in \cite{Patil-etal01} and have been efficiently exploited in AUS as well as in most of the ensemble prediction and data assimilation applications, as for instance the Local Ensemble Kalman Filter (\cite{Ott04,Szunyogh05,Corazza07}) and the Local Ensemble Transform Kalman Filter (\cite{Hunt07}). 

\subsection{Experiments with noisy observations}

Although very instructive and useful in providing insight into possible advantages and drawbacks of a given algorithm, the use of perfect observation is not realistic: in real applications, the assimilation of noisy observations introduces unstructured random errors.
Therefore, we now turn to noisy observations experiments; observations are generated according to the procedure described in Sect.~\ref{Model}. A set of observation realizations, Eq. (\ref{EnKF-obs}), is used to update the ensemble forecast in the EnKF by means of Eq. (\ref{EnKF-analysis}), while Eq. (\ref{AUS-ana-N1-3}) is used for the AUS assimilation in the 3DVar-AUS experiment.

\begin{figure*}[t]
\vspace*{2mm}
\begin{center}
\includegraphics[width=8.3cm]{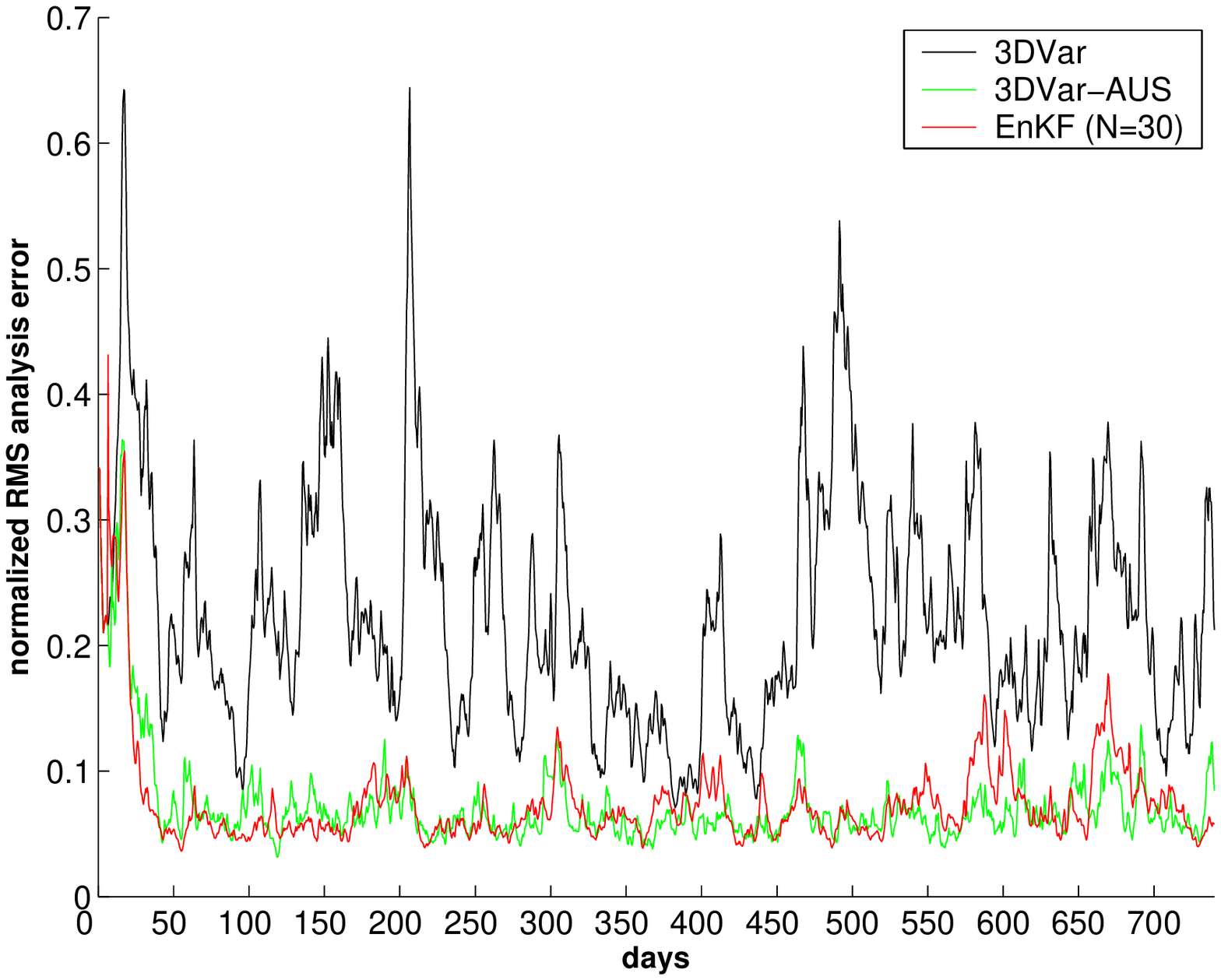}
\end{center}
\caption{\label{ERR-noisy} Noisy observations: normalized RMS analysis error as a function of time for 3DVar (black), 3DVar-AUS (green) and the EnKF (red). Errors are normalized with natural variability and expressed with potential enstrophy norm.} 
\end{figure*}

Figure \ref{ERR-noisy} shows the normalized RMS analysis error, as a function of time, relative to the 2 years trajectory obtained by the assimilation of noisy observations with 3DVar, 3DVar-AUS and EnKF. 
The 3DVar RMS error, excluding the first 100 days, is now about $22\textrm{ }\%$ of the system's natural variability, more than twice the RMS observation error, and large spikes are still present. Remarkably, both 3DVar-AUS and EnKF have an average RMS analysis error of about $7\textrm{ }\%$ of the natural system's variability, below the RMS observational error ($10\textrm{ }\%$, Sect.~\ref{Model}).

Reducing the number of ensemble members to $25$ and $20$ (optimized values $d_0=3000$ Km and $\alpha=0.09$ in both cases), the EnKF RMS analysis error is equal to $10.08\textrm{ }\%$ and $14.21\textrm{ }\%$ respectively.

Notice that besides having comparable time mean values, the 3DVar-AUS and EnKF RMS analysis error shows highly correlated fluctuations which can be interpreted as induced by the same flow-dependent instabilities of the underlying dynamics.
The large improvement of the 3DVar-AUS over the standard 3DVar appears impressive considering that a single BDAS mode and very few observations are used by AUS at each assimilation time.

During the 2 years of simulated time, the AUS analysis was performed $89\textrm{ }\%$ of the total assimilation times: the analysis update was based, on average, on $\hat{N}^{o}=9$ observable structures $\hat{{\bf e}}^{o}_i$ each one detected by (most of the times) $M_i=1$ observation.

These findings confirm the authors' claim that a limited number of observations and unstable directions, if properly used, allow to improve upon the 3DVar performance and make the hybrid scheme performance comparable to that of the EnKF.

Figures 6 and 8 show the 3DVar (left column) and the EnKF (right column) analysis increment superimposed to the 12 hrs actual forecast error, in a sequence of three assimilation times starting at day 204, 00UTC. For the same times, Figs. 7 and 9 show, for the 3DVar-AUS experiment, the increment obtained by AUS only (left column) and 3DVar (right column) superimposed to the 12 hours forecast error and to the background error after the AUS update.

In all panels, black dots indicate the locations of the observations used by the assimilation algorithm (at the considered assimilation time). Figures 6 and 8 show the top PT while the mid level PV is shown in Figs. 7 and 9. 

The 3DVar increments have the typical almost isotropic features coming from the simplified assumption in the definition of the stationary matrix ${\bf B}$.
 The EnKF increments, on the other hand, reproduce very well the features present in the forecast error: by making use of all the available observations and exploiting the accurate description of the actual forecast error realized by the 30 members ensemble-based ${\bf P}^f_{EnKF}$, the EnKF provides reliable analysis updates.

The 3DVar-AUS increments reveal a number of remarkable features. From Fig. 7, we see that the number of observations actually used by AUS is very small; 15 scalar observations only are used at day 204, 00UTC and 9 observations only at days 204, 12UTC and 205, 00UTC. 
For most of the prominent local maxima in forecast error 
the AUS analysis increment has the proper spatial structure that leads to their reduction. 
At day 204, 00UTC, for instance, large error spikes are present along the mid latitude of the channel. Two areas are clearly identifiable: one, north-south oriented, between the 5th and the 10th longitudinal gridpoints and another, located between the 35th and the 55th longitudinal gridpoints. 
Both areas are characterized by positive and negative error elongated structures. The AUS analysis correction accurately reproduces most of these structures; just a small number of observations is sufficient to obtain a reliable and dynamically consistent analysis. 
For instance, the elliptic-shaped, westernmost maximum, located between longitudinal gridpoints 5 and 10, is accurately accounted for by the localized structure extracted from the BDAS mode. However the figure indicates that although the analysis increment has the proper shape, residual background error is present as shown in the corresponding right panel. Analogously, the strong error maximum at the east side of the channel, between longitudinal gridpoints 45 and 55, is reproduced by the AUS increment and partially reduced by using a single observation only. The figure also shows a number of other forecast error local maxima, some of which captured by AUS. This is the case, for instance, of two small dipole-type corrections located between 10 - 15 longitude, 18 - 21 latitude and between 40 - 45 longitude, 3 - 9 latitude, each obtained by assimilating two observations.  

The need to use the 3DVar after the AUS analysis is highlighted by the right panels of the figure. In fact by exploiting the remaining observations 3DVar further reduces the error throughout the domain, and not only in the localized areas of large error development where AUS is most effective.   

After 12 hours, at day 204, 12UTC, there is no evidence of the large error structure which was present in the west side of the channel at the previous analysis time since the AUS assimilation efficiently reduced the forecast error in that area, inhibiting further error growth. 

The error maximum located between 12 - 20 longitude, 6 - 16 latitude, "ignored" by the assimilation, is still present in the forecast error field.

The presence of the strong positive error maximum located, as at the previous analysis time, in the eastern side of the channel, at mid latitude, suggests that the analysis correction at day 204, 00UTC, was not sufficient to eliminate it. Now, at day 204, 12UTC, although its shape is again well captured by the current BDAS mode, the AUS analysis is able only to reduce it. Notice that, a very similar pattern (with opposite sign), manifestation of the same instability, is present in the EnKF forecast error. Similar considerations apply to the last instant shown, day 205, 00UTC.

For all three algorithms, the mid-level PV increments shown in Figs. 8 and 9 present features similar to the PT analysis increments of Figs. 6 and 7. 
It is most remarkable, however, that, except for the single observation at day 204, 12UTC, no observations are used by AUS to provide the analysis update at this level. Still, the analysis increments strongly correlate with the forecast errors, but the analysis is based on observations located at the top level: it is only due to the vertical correlation between the BDAS mode and the forecast error that the analysis update is accurate also at levels different from the observed ones. The single observation at day 204, 12UTC detects the error maximum on the northern side of the channel, between longitudinal gridpoints 21 and 26. 

Figures 7 and 9 illustrate the basic mechanism and the practical conditions necessary for the success of AUS: the ability of the observational network, in terms of its spatial and temporal distribution, to efficiently detect the unstable structures that grow along the trajectory of the assimilation system. They also show the successful use of just a single BDAS mode and a small number of observations at each assimilation time.
The 3DVar analysis update, in this hybrid scheme, allows for the use of all the remaining observations, distributed throughout the domain to correct errors in regions where AUS was not effective. 

The relative contribution of AUS and 3DVAR in the error reduction in the hybrid scheme, is illustrated by Fig. 10. The RMS forecast error, AUS analysis error ({\it i.e.} the background error for 3DVar) and final 3DVar-AUS analysis error, are shown as a function of time, for a 50 days period starting at day 175. The plot shows that AUS and 3DVar give a comparable contribution to error reduction.
This gives further evidence that when instabilities are present the few observations used by AUS (approximately 1\% of the total available observations) are almost as effective as the remaining ones.    
In any case, while the 3DVar always reduces the corresponding background error of approximately the same amount, the effect of AUS is more intermittent.

Finally, Fig. 11 shows the time and domain averaged analysis and forecast errors as a function of the forecast lead time. 
These are all deterministic forecasts; in particular the EnKF forecast is initialized from the EnKF analysis, mean among analysis ensemble members. It is interesting to note that although the EnKF and 3DVar-AUS average analysis error is comparable, the 12 hours forecast error growth is much more rapid in the former case, while it has approximately the same rate afterward.

With the specific setup choices adopted here, the computing time required to complete the 2-year 3DVar-AUS assimilation experiment described in the text, was about $60\textrm{ }\%$ of the time required by the EnKF.

\section{Summary and Discussion}

\subsection{Summary}

A combined data assimilation scheme, the 3DVar-AUS, based on 3DVar and on the dynamically based AUS algorithm, was presented and discussed here. The test ground of the study was the implementation of observation system simulation experiments with a synoptic-like network of observations in the context of an atmospheric quasigeostrophic model. The performance of this proposed scheme was compared with that of the advanced ensemble Kalman filter algorithm. 

According to the formulation given here, 3DVar-AUS is a two-step scheme in which the analysis obtained with AUS is used as the background for the 3DVar analysis update. At each analysis time, an automatic procedure searched for observable structures, extracted from the current BDAS mode, on which the analysis update is based. 

In all the experiments described in the text, just a single global BDAS mode and only a very small number of observations were used by AUS at each assimilation time. 

%{\bf RevA-sc2a}

The system at hand has 24 positive Lyapunov exponents, corresponding to 24 independent diverging directions and inspection of the full spectrum shows that these directions have competitive growth rates \cite{CGTU}. These competitive instabilities are present simultaneously in the single global BDAS mode used at a particular analysis time under the form of localized structures dominated by local instabilities; these are the structures exploited in the AUS assimilation.

%A limited number of scalar observations was actually used by AUS to update the state's estimate: results indicate that, on average, AUS made use of nine observable structures $\hat{{\bf e}}^{o}$ and of a single scalar observation for each of them, at each assimilation time. Anyhow, in spite of the use of such a reduced number of observations, by combining AUS with 3DVar a dramatic improvement was achieved with respect to the standard 3DVar, supporting the claim that few observations, if properly selected and used with the goal of controlling the data assimilation system instabilities, can lead to significant improvements of the overall performance of the assimilation scheme.

A key ingredient for the success of AUS is clearly the accuracy of the BDAS modes to explain relevant components of the actual forecast error. The BDAS procedure is built to estimate the instabilities growing along the complete analysis-forecast cycle and implicitly embeds all the information about the observational forcing. As indicated by Fig.~\ref{BDAS-perf} and, indirectly through the analysis increment in Figs. 6 and 7, the correlation between the single BDAS mode and the actual forecast error is evident. In particular, the BDAS modes reliably reproduce most of the small scales features of the error spatial patterns.

In the model and observational context used here, the EnKF and the 3DVar-AUS showed a comparable performance in terms of accuracy of the state estimate, either with perfect or noisy observations. In the perfect observations case, both algorithms are able to reduce errors to a very low level demonstrating, on the one hand, an efficient tracking of the instabilities (AUS), and, on the other hand, a reliable time-dependent forecast error covariance description (EnKF). With noisy observations, the analysis error of 3DVar-AUS and of EnKF
rapidly decreases, then remains below the observational noise level:
the improvement over the regular 3DVar alone is dramatic in both cases. 

It is remarkable that, despite the fact that 3DVar-AUS and the EnKF attain a comparable average analysis error ($6.8\textrm{ }\%$ and $6.6\textrm{ }\%$ of the system's natural variability, respectively) the 12 hours forecast error in the EnKF is larger than in the 3DVar-AUS ($9.9\textrm{ }\%$ and $7.3\textrm{ }\%$, respectively) and its 24 hours forecast skill is somewhere between the 36 hours and the 48 hours forecast skill of the 3DVar-AUS..

This result confirms the efficiency of AUS in reducing the flow instabilities responsible for the error growth even in a fixed observations setting.

\subsection{Conclusions}

The results of the present study lead to the following interpretation: the forecast error shows very intense and localized structures that are similar in the three assimilation experiments and in particular between the 3DVar-AUS and the EnKF: accordingly, a high correlation in the time evolution of the RMS error of the two schemes is observed. As indicated by these similarities, the local maxima of the forecast error are presumably concentrated in regions where dynamical instabilities, compatible with the observational forcing, develop along the analysis-forecast cycle.

With a fixed observational network, observations are assimilated by AUS only when they are found in correspondence to the local maxima of the unstable structures. Therefore it is necessary to use 3DVar to assimilate the remaining observations.
Three dimensional error structures, provide the corresponding AUS analysis correction, whose amplitude and sign are determined from the innovation: hence its ability to reduce the forecast error also at levels other than the observed ones.

In real world applications, the employment of AUS is expected to give an improvement over a 3DVar (or Optimal Interpolation) every time an unstable structure is observed by either fixed or adaptive observations. When the latter are available and there is no need to wait for the instability to travel into observed regions, the expected improvement in using AUS is enhanced.

Also the EnKF takes advantage of the observations that happen to be (or are intentionally located) in regions where instabilities develop and the present fixed observations experiments prove that the observations are exploited by the two methods with similar efficiency. The EnKF obtains the same goal of capturing and controlling the instabilities, the different approach of the two methods being exemplified by the following idealized example where a single unstable structure is present in the forecast error. In this simple case, the EnKF would identify the unstable structure as difference of the members (two members would be sufficient for this purpose) from the ensemble mean. To estimate the amplitude of the correction, however, a sufficiently large number of representative members would still be necessary to provide a reliable estimate of the associated error variance. In contrast, AUS estimates the direction as difference of a single unstable perturbation from the control and estimates the amplitude of the correction from innovation.

The present result, where EnKF and 3DVar-AUS have a similar analysis performance, seems to indicate that both methods were able to efficiently exploit all the available observations useful to control the flow-dependent instabilities. A point in favor of the 3DVar-AUS scheme is its better performance in the forecast stage.  

Other hybrid schemes have been proposed which combine 3DVar with the ensemble based filters (\cite{HamSny00,EtherBis04,Wang_etal_07a,Wang_etal_07b}).

Apart from the differences among the different approaches proposed in previous studies (see \cite{Wang_etal_07b} and references therein) two results deserve to be mentioned here. Usually, in these hybrid schemes, the forecast error covariance matrix is obtained by combining a static covariance with an ensemble based one, through a tunable scalar weight. Improvements over a 3DVar conveyed by the flow-dependent covariance are less important when a dense observational network is available (\cite{HamSny00}). Moreover the presence of model error significantly reduces the optimal value of the weight to be given to the flow-dependent covariance (\cite{EtherBis04}). 

Based on these works and on the results of the present study we draw the following conclusion.

To hybridize an ensemble based scheme or AUS with a static covariance might turn to be more convenient when instabilities loose some of their significance, either because the observational network is particularly dense (\cite{Whitaker_et_al_2007}) or if the model error is such that the model does not provide an adequate representation of the true unstable modes (\cite{EtherBis04}).
In view of the latter consideration, while the hybrid approach proposed in the present study may turn out to be useful also in the presence of model error, further work is needed to address this problem.

Because the performance of different schemes depends upon many different factors, including model error and the number and distribution of observations, it remains to be seen which method is more efficient in a particular operational setting. Therefore a quantitative comparison relevant for operational purposes is out of the scope of the present study.

However, given the ubiquity of the role of instabilities in degrading the analysis accuracy, we believe that AUS may turn out to be useful in those circumstances in which the observational network, and in particular an adaptive one, allows their detection.  

\section*{Acknowledgements}
This work was supported by the HPC-Europa project (RII3-CT-2003-506079) of the European Community - Research Infrastructure Action under the FP6 "Structuring the European Research Area" Program, and by Belgian Federal Science Policy Program under contract MO/34/017.

\section*{Appendix - Estimate of the forecast error variance along the BDAS mode}    %% Appendix B

In the AUS analysis update Eq. (\ref{AUS-ana-N1-3}), each of the scalars $\gamma_i^2$, representing the variance of the forecast error along the structure $\hat{{\bf e}}_i^{o}$, need to be estimated. Following the approach used in \cite{CTU} this is done statistically, by making use of the innovations.  

In practice the terms $\gamma_i^{2}({\bf H}_i\hat{{\bf e}}_i^{o})^T({\bf H}_i\hat{{\bf e}}_i^{o})$ are estimated as follows:
\begin{equation}
\gamma_i^{2}({\bf H}_i\hat{{\bf e}}_i^{o})^T({\bf H}_i\hat{{\bf e}}_i^{o})\approx \left\{\begin{array}{ccc} \frac{1}{D}\langle{\bf d}^T_i,{\bf d}_i\rangle_{T}-\sigma^2, \qquad \frac{1}{D}\langle{\bf d}^T_i,{\bf d}_i\rangle_{T} > \sigma^2 \\
\\ \frac{1}{D}\langle{\bf d}^T_i,{\bf d}_i\rangle_{T}, \qquad \frac{1}{D}\langle{\bf d}^T_i,{\bf d}_i\rangle_{T} \le \sigma^2 \\ \end{array}\right.
\end{equation}
where $\langle,\rangle_{T}$ represents the average over an appropriate time interval $T$ and $D$ is a scalar coefficient to be tuned. After optimization, these values are set to $D=1.35$ and $T=8$ days.

Based on Eq. (18), if ${\bf He}$ approaches zero the analysis correction approaches infinity. However because only observations in proximity of maxima in the BDAS mode are chosen and in view of the selection procedure described in Sect. 3.3 (point (b)) this circumstance is never encountered.

%\section{\\ \\ \hspace*{-7mm} HEADING}    %% Appendix A

%\subsection                               %% Appendix A1, A2, etc.

%\begin{thebibliography}{}

%\bibitem[AUTHOR(YEAR)]{LABEL}
%REFERENCE 1

%\bibitem[AUTHOR(YEAR)]{LABEL}
%REFERENCE 2

%...
\bibliography{biblio}

\begin{thebibliography}{10}

\bibitem{Anderson03}
J.L. Anderson.
\newblock A local least squares framework for ensemble filtering.
\newblock {\em Mon. Wea. Rev.}, {131}({}):634--642, 2003.

\bibitem{AndersonAnderson99}
J.L. Anderson and S.~Anderson.
\newblock A monte carlo implementation of the nonlinear filtering problem to
  produce ensemble assimilation and forecast.
\newblock {\em Mon. Wea. Rev.}, {127}({}):2741--2758, 1999.

\bibitem{Benetal80}
G.~Benettin, L.~Galgani, A.~Giorgilli, and J.M. Strelcyn.
\newblock Lyapunov characteristic exponents for smooth dynamical systems and
  for hamiltonian systems; a method for computing them.
\newblock {\em Meccanica}, {\bf 15}({}):9--30, 1980.

\bibitem{bergman79}
K.~H. Bergman.
\newblock Multivariate analysis of temperature and winds using optimum
  interpolation.
\newblock {\em Mon. Wea. Rev.}, {107}({ }):1423--1444, 1979.

\bibitem{Bleck78}
R.~Bleck.
\newblock Simulation of coastal upwelling frontogenesis with an isopycnic
  coordinate model.
\newblock {\em J. Geophys. Res.}, 83C:6163--6172, 1978.

\bibitem{CGTU}
A.~Carrassi, M.~Ghil, A.~Trevisan, and F.~Uboldi.
\newblock Data assimilation as a nonlinear dynamical system problem: Stability
  and convergence of the prediction-assimilation system.
\newblock {\em Chaos}, {under review}({ }), 2007.

\bibitem{CTU}
A.~Carrassi, A.~Trevisan, and F.~Uboldi.
\newblock Adaptive observations and assimilation in the unstable subspace by
  breeding on the data-assimilation system.
\newblock {\em Tellus}, {59A}({ }):101--113, 2007.

\bibitem{Corazza03}
M.~Corazza, E.~Kalnay, D.J. Patil, S.-C. Yang, R.~Morss, M.~Cai, I.~Szunyogh,
  B.R. Hunt, and J.A. Yorke.
\newblock Use of the breeding technique to estimate the structure of the
  analysis 'error of the day'.
\newblock {\em Nonlin. Processes Geophys.}, 10:233--243, 2003.

\bibitem{Corazza07}
M.~Corazza, E.~Kalnay, and S.-C. Yang.
\newblock An implementation of the local ensemble kalman filter for a simple
  quasi-geostrophic model: Results and comparison with a 3d-var data
  assimilation system.
\newblock {\em Nonlin. Processes Geophys.}, 14:89--101, 2007.

\bibitem{DescampsTalagrand07}
L.~Descamps and O.~Talagrand.
\newblock On some aspects of the definition of initial conditions for ensemble
  prediction.
\newblock {\em Mon. Wea. Rev.}, accepted({ }), 2007.

\bibitem{EtherBis04}
B.J. Etherton and C.H. Bishop.
\newblock Resilience of hybrid ensemble/3dvar analysis schemes to model error
  and ensemble covariance error.
\newblock {\em Mon. Wea. Rev.}, {\bf 132}:1065--1080, 2004.

\bibitem{Even94}
G.~Evensen.
\newblock Inverse methods and data assimilation in nonlinear ocean models.
\newblock {\em Physica D}, {\bf 77}:108--129, 1994.

\bibitem{Even03}
G.~Evensen.
\newblock The ensemble kalman filter: theoretical formulation and practical
  implementation.
\newblock {\em Oc. Dyn.}, 53:343--367, 2003.

\bibitem{Even04}
G.~Evensen.
\newblock Sampling strategies and square root analysis schemes for the enkf.
\newblock {\em Oc. Dyn.}, 53:539--560, 2004.

\bibitem{GaspariCohn99}
G.~Gaspari and S.~Cohn.
\newblock Construction of correlation functions in two and three dimensions.
\newblock {\em Quart. J. Roy. Meteor. Soc.}, 125({ }):723--757, 1999.

\bibitem{HamSny00}
T.~M. Hamill and C.~Snyder.
\newblock A hybrid ensemble kalman filter 3d variational scheme.
\newblock {\em Mon. Wea. Rev.}, {\bf 129}:2905--2919, 2000.

\bibitem{houtekamer93}
P.~L. Houtekamer.
\newblock Global and local skill forecast.
\newblock {\em Mon. Wea. Rev.}, 121:1834--1846, 1993.

\bibitem{houtekamer01}
P.~L. Houtekamer and H.~L. Mitchell.
\newblock A sequential ensemble kalman filter fot atmospheric data
  assimilation.
\newblock {\em Mon. Wea. Rev.}, 129:123--137, 2001.

\bibitem{Hunt07}
B.R. Hunt, E.~Kostelich, and I.~Szunyogh.
\newblock Efficient data assimilation for spatiotemporal chaos: a local
  ensemble transform kalman filter.
\newblock {\em Physica D}, page in print, 2007.

\bibitem{ide97}
K.~Ide, P.~Courtier, M.~Ghil, and A.~Lorenc.
\newblock Unified notation for data assimilation: Operational, variational and
  sequential.
\newblock {\em J. Met. Soc. Japan}, {\bf 75}:181--189, 1997.

\bibitem{Jaz70}
A.~H. Jazwinski.
\newblock {\em Stochastic Processes and Filtering Theory}.
\newblock Academic Press, 1970.

\bibitem{Langland2005}
R.~H. Langland.
\newblock Observation impact during the north atlantic trec-2003.
\newblock {\em Mon. Wea. Rev.}, {\bf 133}({ }):2297--2309, 2005.

\bibitem{Lorenz96}
E.N. Lorenz.
\newblock {\em Predictability: A problem partly solved.}
\newblock ${\it Proc. Seminar\ on\ Predictability}$ Vol.1, ECMWF, Reading,
  Berkshire, UK, pp. 1-18, 1996.

\bibitem{morss99}
R.~E. Morss.
\newblock {\em Adaptive observations: Idealized sampling strategies for
  improving numerical weather prediction.}
\newblock PhD thesis, Massachusetts Institute of Technology, 1999.

\bibitem{Morss01}
R.E. Morss, K.A. Emanuel, and C.~Snyder.
\newblock Idealized adaptive observation strategies for improving numerical
  weather prediction.
\newblock {\em J. Atmos. Sci.}, 58:210--232, 2001.

\bibitem{Ochotta_etal2005}
T.~Ochotta, C.~Gebhardt, D.~Saupe, and W.~Wergen.
\newblock Adaptive thinning of atmospheric observations in data assimilation
  with vector quantization and filtering methods.
\newblock {\em Quart. J. Roy. Meteorol. Soc.}, 131:3427--3437, 2005.

\bibitem{Ott04}
E.~Ott, B.R. Hunt, I.~Szunyogh, A.V. Zimin, E.J. Kostelich, M.~Corazza,
  E.~Kalnay, D.J. Patil, and J.A. Yorke.
\newblock A local ensemble kalman filter for atmospheric data assimilation.
\newblock {\em Tellus}, 56:415--428, 2004.

\bibitem{Patil-etal01}
D.J. Patil, B.R. Hunt, E.~Kalnay, J.A. Yorke, and E.~Ott.
\newblock Local low dimensionality of atmospheric dynamics.
\newblock {\em Phys. Rev. Lett.}, 86({}):5878--5881, 2001.

\bibitem{RotBao96}
R.~Rotunno and J.W. Bao.
\newblock A case study of cyclogenesis using a model hierarchy.
\newblock {\em Mon. Weather Rev.}, 124({}):1051--1066, 1996.

\bibitem{snyderhamill03}
C.~Snyder and T.~H. Hamill.
\newblock Leading lyapunov vectors of a turbolent baroclinic jet in a
  quasigeostrophic model.
\newblock {\em J. Atmos. Sci.}, 60({}):683--688, 2003.

\bibitem{Szunyogh05}
I.~Szunyogh, E.J. Kostelich, G.~Gyarmati, D.J. Patil, E.~Kalnay, E.~Ott, and
  J.~Yorke.
\newblock Assessing a local ensemble kalman filter: Perfect model experiments
  with the national center for the environmental prediction global model.
\newblock {\em Tellus}, 57({}):528--545, 2005.

\bibitem{Szunyogh02}
I.~Szunyogh, Z.~Toth, A.V. Zimin, S.J. Majumdar, and A.~Persson.
\newblock Propagation of the effect of targeted observations: The 2000 winter
  storm reconnaissance program.
\newblock {\em Mon. Wea. Rev.}, 130({}):1144--1165, 2002.

\bibitem{talagrand97}
O.~Talagrand.
\newblock Assimilation of observations, an introduction.
\newblock {\em J. Met. Soc. Japan}, 75({1B}):191--209, 1997.

\bibitem{Tippetetal2003}
M.K. Tippet, J.L. Anderson, C.H. Bishop, T.H. Hamill, and J.S. Whitaker.
\newblock Ensemble square root filters.
\newblock {\em Mon. Wea. Rev.}, {\bf 131}:1485--1490, 2003.

\bibitem{TothKalnay93}
Z.~Toth and E.~Kalnay.
\newblock Ensemble forecasting at nmc. the generation of perturbations.
\newblock {\em Bull. Amer. Meteor. Soc.}, 74({}):2317--2330, 1993.

\bibitem{TothKalnay97}
Z.~Toth and E.~Kalnay.
\newblock Ensemble forecasting at ncep: the breeding method.
\newblock {\em Mon. Wea. Rev.}, 125({}):3297--3318, 1997.

\bibitem{TrevPan98}
A.~Trevisan and F.~Pancotti.
\newblock Periodic orbits, lyapunov vectors and singular vectors in the lorenz
  system.
\newblock {\em J. Atmos. Sci.}, {\bf 55}({}):390--398, 1998.

\bibitem{TU}
A.~Trevisan and F.~Uboldi.
\newblock Assimilation of standard and targeted observations within the
  unstable subspace of the observation-analysis-forecast cycle system.
\newblock {\em J. Atmos. Sci.}, 61({}):103--113, 2004.

\bibitem{UT}
F.~Uboldi and A.~Trevisan.
\newblock Detecting unstable structures and controlling error growth by
  assimilation of standard and adaptive observations in a primitive equation
  ocean model.
\newblock {\em Nonlin. Processes Geophys.}, 13({}):67--81, 2006.

\bibitem{UTC}
F.~Uboldi, A.~Trevisan, and A.~Carrassi.
\newblock Developing a dynamically based assimilation method for targeted and
  standard observations.
\newblock {\em Nonlin. Processes Geophys.}, 12({}):149--156, 2005.

\bibitem{Wang_etal_07a}
X.~Wang, T.H. Hamill, J.S. Whitaker, and C.H. Bishop.
\newblock A comparison of hybrid ensemble transform kalman filter-oi and
  ensemble square root filter analysis schemes.
\newblock {\em Mon. Wea. Rev.}, {\bf 135}:1055--1076, 2007.

\bibitem{Wang_etal_07b}
X.~Wang, C.~Snyder, and T.H. Hamill.
\newblock On the theoretical equivalence of differently proposed ensemble-3dvar
  hybrid analysis schemes.
\newblock {\em Mon. Wea. Rev.}, {\bf 135}:222--227, 2007.

\bibitem{WhitaHam02}
J.~Whitaker and T.~Hamill.
\newblock Ensemble data assimilation without perturbed observations.
\newblock {\em Mon. Wea. Rev.}, {\bf 130}:1913--1924, 2002.

\bibitem{Whitaker_et_al_2007}
J.~Whitaker, T.M. Hamill, X.~Wei, Y.~Song, and Z.~Toth.
\newblock Ensemble data assimilation with the ncep global forecast system.
\newblock {\em Mon. Wea. Rev.}, {\bf under revision}, 2007.

\bibitem{WolfeSamel07}
C.~L. Wolfe and R.~M. Samelson.
\newblock An efficient method for recovering lyapunov vectors from singular
  vectors.
\newblock {\em Tellus}, {\bf 59A}:355--366, 2007.

\end{thebibliography}
\bibliographystyle{plain}
\newpage
%\bibliographystyle{copernicus}
%\end{thebibliography}

%% Literature citations
%% command                        & example result
%% \citet{jones90}|               & Jones et al.\ (1990)
%% \citep{jones90}|               & (Jones et al., 1990)
%% \citep{jones90,jones93}|       & (Jones et al., 1990, 1993)
%% \citep[p.~32]{jones90}|        & (Jones et al., 1990, p.~32)
%% \citep[e.g.,][]{jones90}|      & (e.g., Jones et al., 1990)
%% \citep[e.g.,][p.~32]{jones90}| & (e.g., Jones et al., 1990, p.~32)
%% \citeauthor{jones90}|          & Jones et al.
%% \citeyear{jones90}|            & 1990

%% FIGURES %%%%%%%%%%%%%%%%%%%%%%%%%%%%%%%%%%%%%%%%%%%%%%%%%%%%%%%%%%%%%%%%%%%%

%% ONE-COLUMN FIGURES

%f

%% TWO-COLUMN FIGURES

%f

\begin{figure*}[t]
%\addtocounter{figure}{0}
%\renewcommand{\thefigure}{\arabic{figure}a}
\vspace*{2mm}
\begin{center}
\includegraphics[width=17cm]{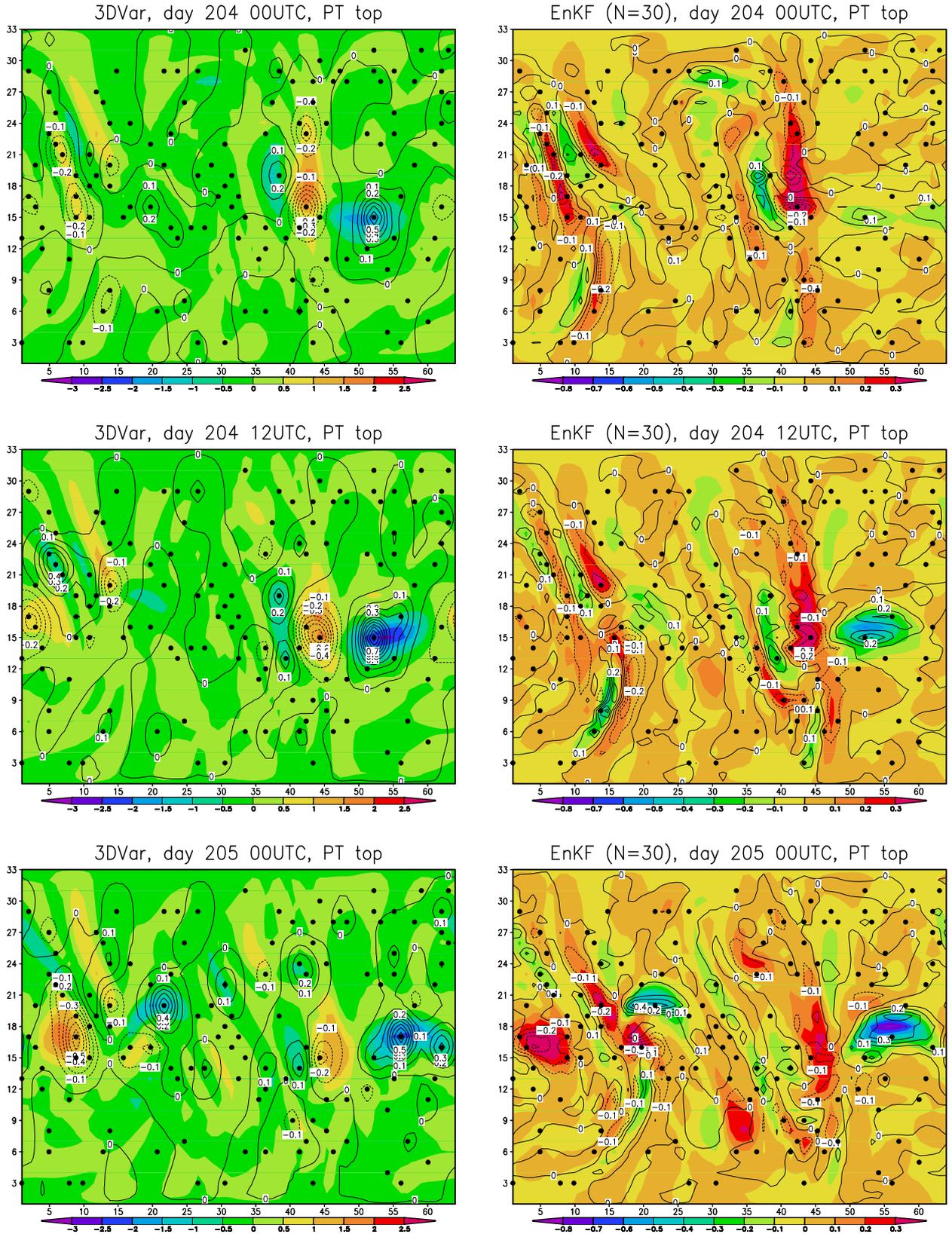}
\caption{Noisy observations experiments: top level potential temperature (PT) 12hr forecast error (shaded) and analysis increment (contour) in a sequence of three assimilation times starting at day 204, for 3DVar (left column) and EnKF (right column). Black dots indicate locations of the observations used.} 
\end{center}
\end{figure*}

\begin{figure*}[t]
%\addtocounter{figure}{-1}
%\renewcommand{\thefigure}{\arabic{figure}b}
\vspace*{2mm}
\begin{center}
\includegraphics[height=16cm, width=17cm]{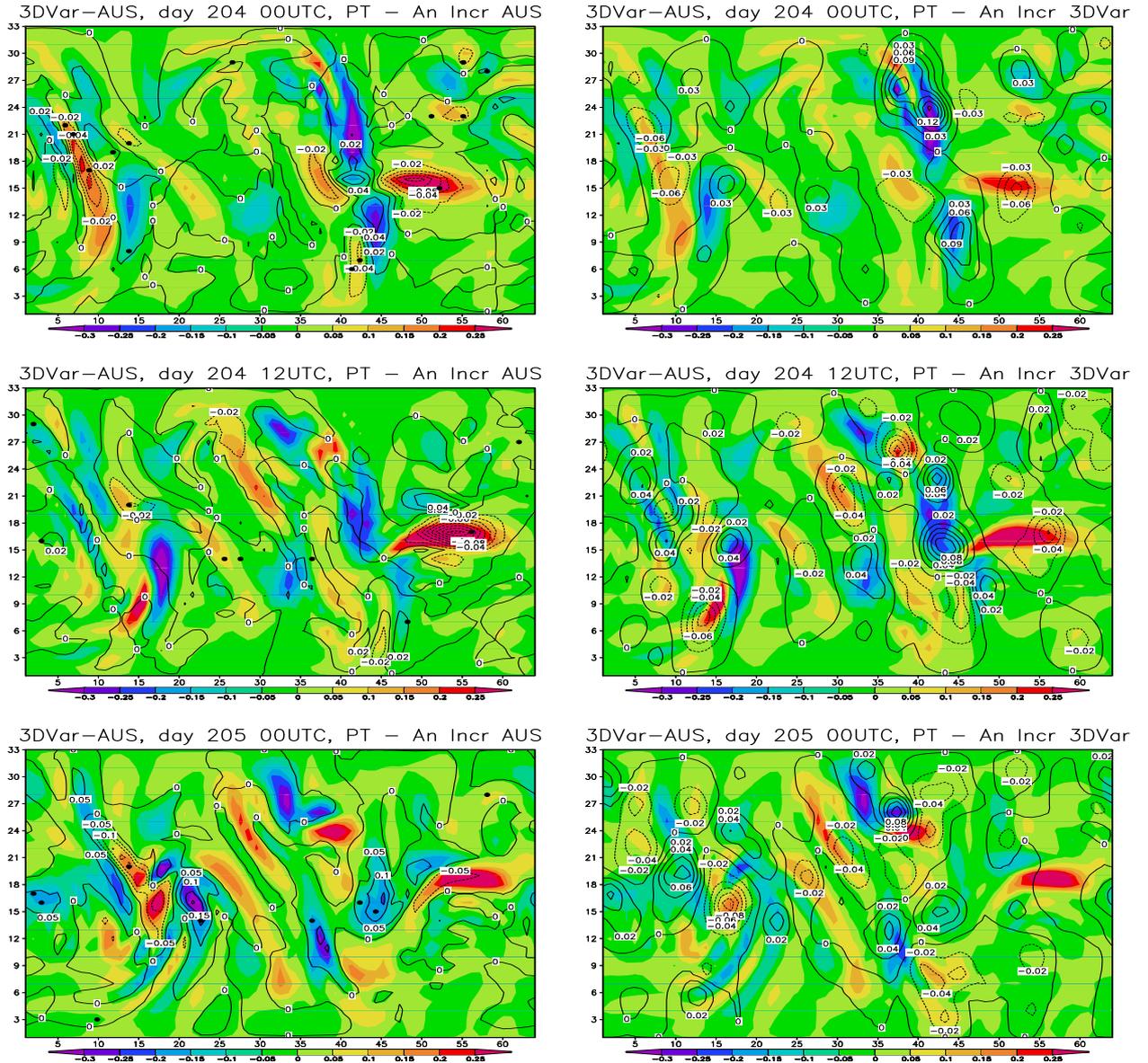}
\end{center}
\caption{ Noisy observations experiments: top level potential temperature (PT) 12hr forecast error (shaded) and analysis increment (contour) in a sequence of three assimilation times starting at day 204, for 3DVar-AUS. The rights panels show the AUS analysis increment superimposed to the 12 hours forecast error; the left panels show the 3DVar analysis increment superimposed to the AUS analysis error (which is used as the background in the 3DVar analysis update). Black dots indicate locations of the observations used.} 
\label{FIG6b}
\end{figure*}

\begin{figure*}[t]
%\addtocounter{figure}{0}
%\renewcommand{\thefigure}{\arabic{figure}a}
\vspace*{2mm}
\begin{center}
\includegraphics[width=17cm]{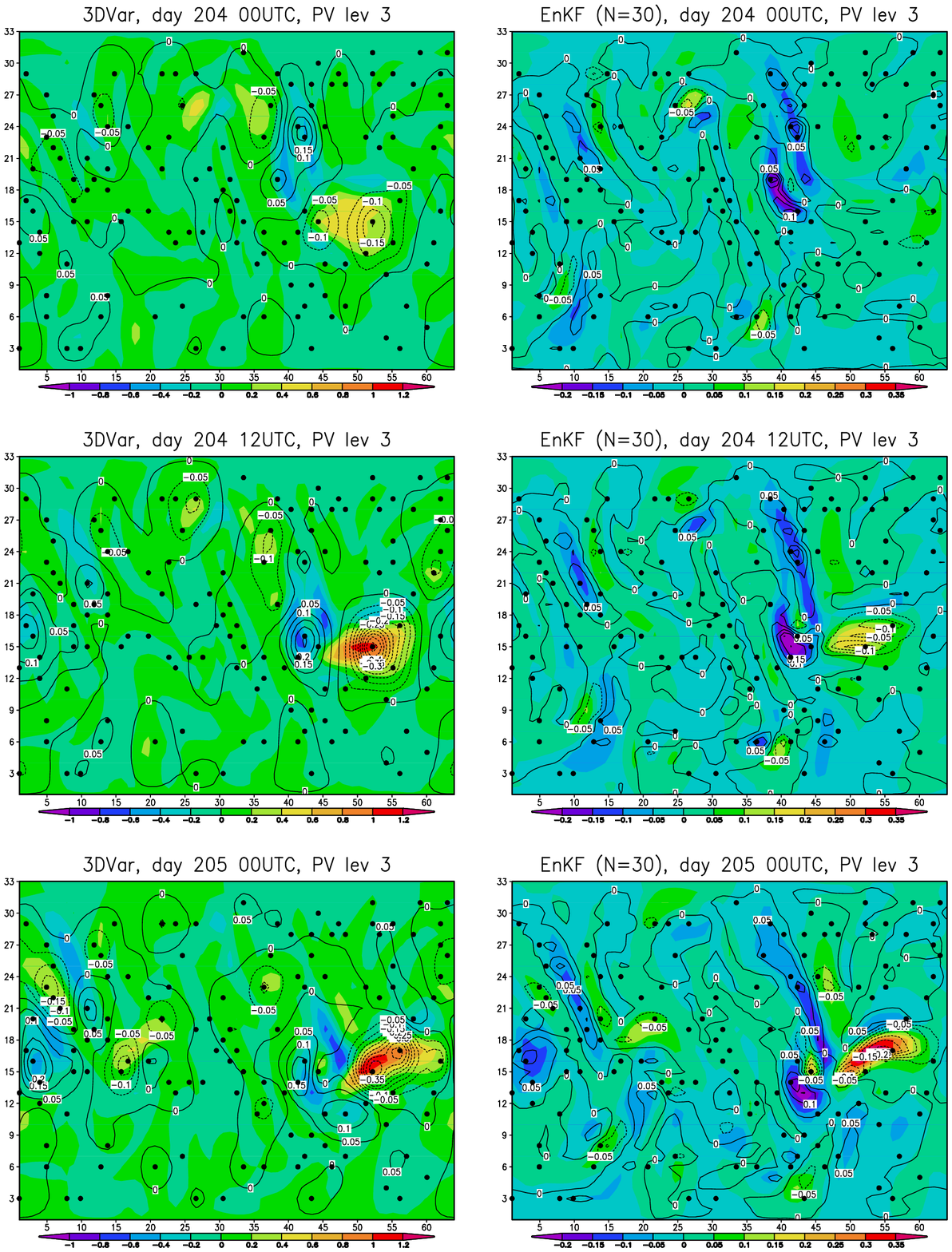}
\end{center}
\caption{ The same as Fig.6 but the mid level potential vorticity (PV) field is shown. }
\end{figure*}

\begin{figure*}[t]
%\addtocounter{figure}{-1}
%\renewcommand{\thefigure}{\arabic{figure}b}
\vspace*{2mm}
\begin{center}
\includegraphics[width=17cm]{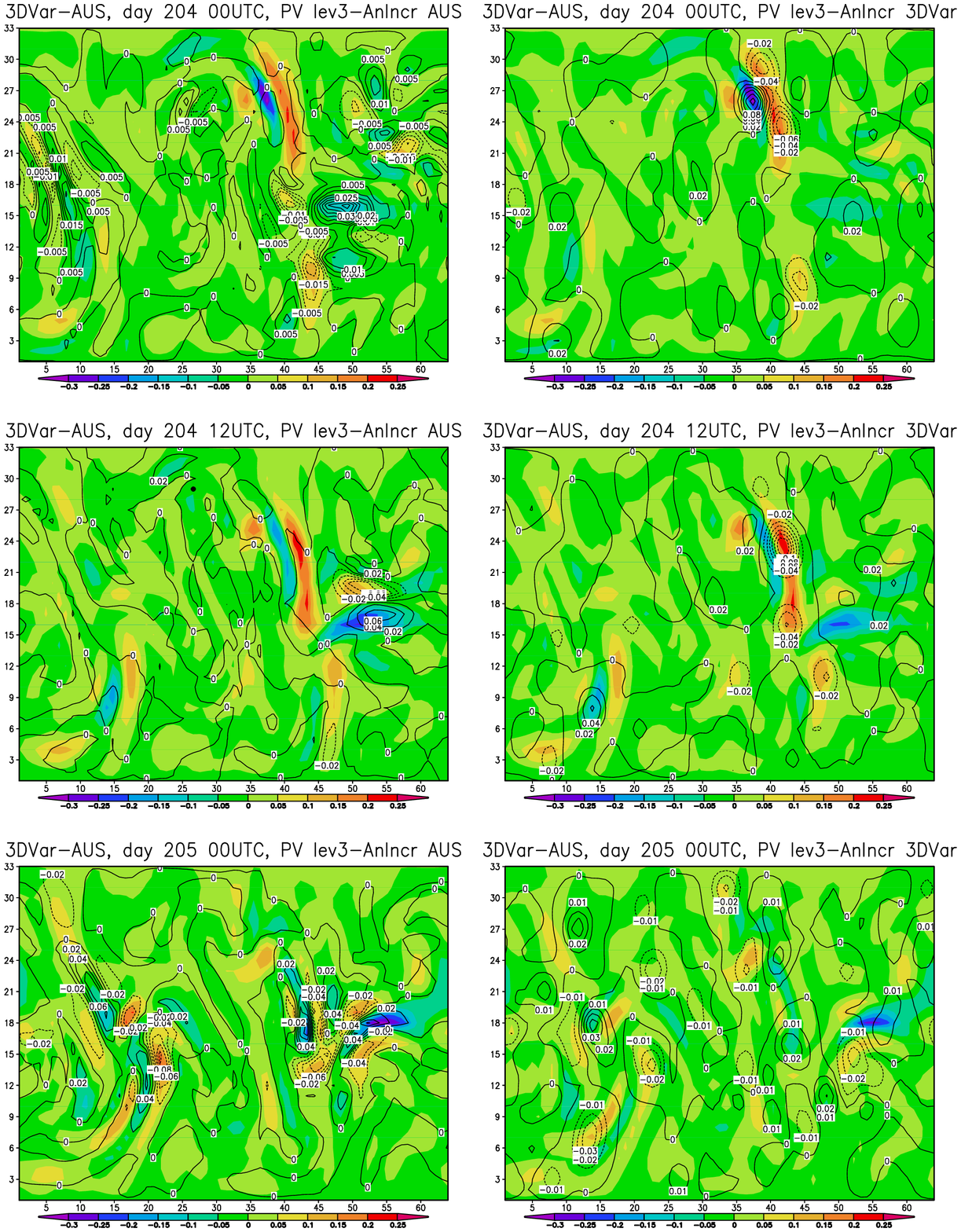}
\end{center}
\caption{ The same as Fig.7 but the mid level potential vorticity (PV) field is shown. }
\end{figure*}

\begin{figure*}[t]
\vspace*{2mm}
\begin{center}
\includegraphics[width=8.3cm]{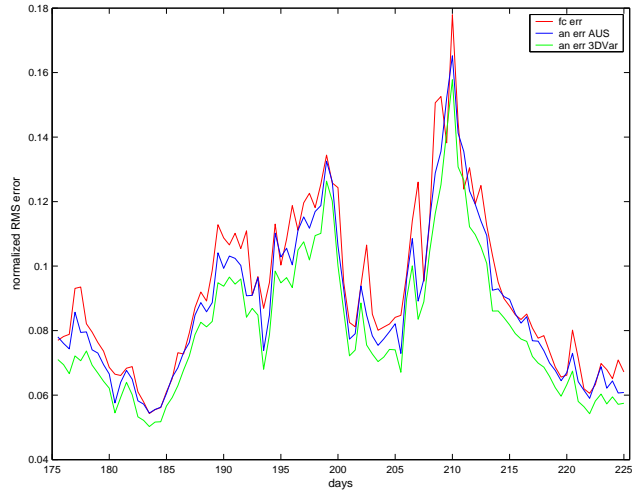}
\end{center}
\caption{Normalized RMS analysis error as a function of time for the 3DVar-AUS experiments. Noisy observations.} 
\end{figure*}

\begin{figure*}[t]
\vspace*{2mm}
\begin{center}
\includegraphics[width=8.3cm]{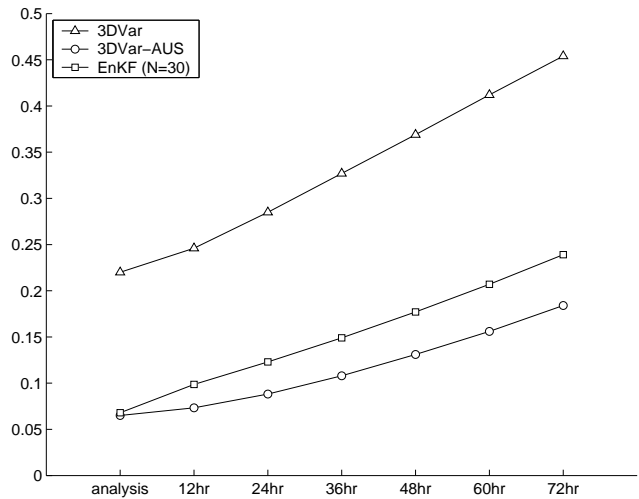}
\end{center}
\caption{\label{FCerr}Noisy observations experiment. Time and domain normalized RMS analysis and forecast error as a function of the forecast range: 3DVar (triangles), 3DVar-AUS (circles), EnKF (squares).}
\end{figure*}

%\printfigures
%% TABLES %%%%%%%%%%%%%%%%%%%%%%%%%%%%%%%%%%%%%%%%%%%%%%%%%%%%%%%%%%%%%%%%%%%%

%% ONE-COLUMN TABLE

%t
%\begin{table}[t]
%\caption{TEXT}
%\vskip4mm
%\centering
%\begin{tabular}{column = lcr}
%\tophline

%\middlehline

%\bottomhline
%\end{tabular}
%\end{table}

%% TWO-COLUMN TABLE

%t
%\begin{table*}[t]
%\caption{TEXT}
%\vskip4mm
%\centering
%\begin{tabular}{column = lcr}
%\tophline

%\middlehline

%\bottomhline
%\end{tabular}
%\end{table*}

%% The different columns must be seperated with a & command and should
%% end with \\ to identify the column brake.

%%%%%%%%%%%%%%%%%%%%%%%%%%%%%%%%%%%%%%%%%%%%%%%%%%%%%%%%%%%%%%%%%%%%%%%%%%%%%%

%% If figures and tables must be numbered 1a, 1b, etc. the following command
%% should be inserted before the begin{} command.

%\addtocounter{figure}{-1}\renewcommand{\thefigure}{\arabic{figure}a}

\end{document}